\RequirePackage{fix-cm}
\RequirePackage{amsmath}

\documentclass[10pt,a4paper]{article}
\usepackage{graphicx}
\usepackage{amsfonts}
\usepackage{amssymb}
\usepackage{bm}
\usepackage{hyperref}
\usepackage{multirow}
\usepackage{tabularx}
\usepackage{xcolor}
\usepackage{tikz}
\usetikzlibrary{calc,positioning}

\usepackage[author={Ulm}]{pdfcomment}

\newcolumntype{V}[1]{>{\centering\arraybackslash} m{#1} }
\newcolumntype{C}[1]{>{\centering\let\newline\\\arraybackslash\hspace{0pt}} m{#1} }

\newcommand{\R}{\mathbb{R}}
\newcommand{\N}{\mathbb{N}}

\newcommand{\x}{\mathbf{x}}
\newcommand{\bN}{\mathbf{N}}

\newcommand{\VNR}{\operatorname{VNR}}
\newcommand{\nof}{\operatorname{nof}}
\newcommand{\vol}{\operatorname{vol}}

\newcommand{\dsc}{\operatorname{dsc}}
\newcommand{\bary}{\operatorname{bar}}

\newcommand{\card}{\operatorname{card}}

\providecommand{\keywords}[1]{\noindent\textit{Keywords: } #1}

\usepackage[]{footmisc}
\usepackage{titlesec}

\titleformat{\section}
  {\normalfont\fontsize{12}{15}\bfseries}{\thesection}{1em}{}

\titleformat{\subsection}
  {\normalfont\fontsize{12}{15}\itshape}{\thesubsection}{1em}{}

\titleformat{\subsubsection}
  {\normalfont\fontsize{10}{15}\itshape}{\thesubsubsection}{1em}{}

\begin{document}

\title{Exploration of Gibbs-Laguerre tessellations for three-dimensional stochastic modeling
}


\author{F. Seitl\footnote{tel.: 00420951553327, e-mail: seitl@karlin.mff.cuni.cz}\footnote{Charles University Prague, Faculty of Mathematics and Physics, Department of Probability and Mathematical Statistics, Sokolovsk\'{a} 83, 186 75 Praha 8} , L. Petrich\footnote{ Ulm University, Faculty of Mathematics and Economics, Institute of Stochastics, 89069 Ulm}, J. Stan\v{e}k\footnote{Charles University Prague, Faculty of Mathematics and Physics, Department of Mathematics Education, Sokolovsk\'{a} 83, 186 75 Praha 8}, \\ C.E. Krill III\footnote{Ulm University, Faculty of Engineering, Computer Science and Psychology, Institute of Functional Nanosystems, 89069 Ulm}, V. Schmidt$^3$,  V. Bene\v{s}$^2$
}


\date{}


\begin{center}
\LARGE{Exploration of Gibbs-Laguerre tessellations for three-dimensional stochastic modeling}

\vspace{1.2cm}

\large{F. Seitl\footnote[1]{tel.: 00420951553327, e-mail: seitl@karlin.mff.cuni.cz}\footnote[2]{\label{fn2}Charles University Prague, Faculty of Mathematics and Physics, Department of Probability and Mathematical Statistics, Sokolovsk\'{a} 83, 186 75 Praha 8} , L. Petrich\footnote[3]{\label{fn3} Ulm University, Faculty of Mathematics and Economics, Institute of Stochastics, 89069 Ulm}, J. Stan\v{e}k\footnote[4]{Charles University Prague, Faculty of Mathematics and Physics, Department of Mathematics Education, Sokolovsk\'{a} 83, 186 75 Praha 8}, \\ C.E. Krill III\footnote[5]{Ulm University, Faculty of Engineering, Computer Science and Psychology, Institute of Functional Nanosystems, 89069 Ulm}, V. Schmidt\footnotemark[3], V. Bene\v{s}\footnotemark[2]{}
}
\end{center}

\normalsize

\vspace{0,7cm}

\begin{abstract}
Random tessellations are well suited for probabilistic modeling of three-dimensional (3D) grain microstructures of polycrystalline materials. The present paper is focused on so-called Gibbs-Laguerre tessellations, in which the generators of the Laguerre tessellation form a Gibbs point process. The goal is to construct an energy function of the Gibbs point process such that the resulting tessellation matches some desired geometrical properties. Since the model is analytically intractable, our main tool of analysis is stochastic simulation based on Markov chain Monte Carlo. Such simulations enable us to investigate the properties of the models, and, in the next step, to apply the knowledge gained to the statistical reconstruction of the 3D microstructure of an aluminum alloy extracted from 3D tomographic image data.

\medskip

\end{abstract}

\section{Introduction}
\label{intro}

In materials science, discovering and quantifying relationships between microstructure and bulk properties of materials is one of the most important research goals \cite{GR}. The traditional approach is to analyze samples of real materials. Whereas this method arguably returns the most realistic results, it is time consuming and demanding to produce, image and investigate the specimens. With the increase in readily available computing power, it is possible to support such investigations today with \textit{in silico} experiments, which drastically reduce the time spent in the lab, see, e.g., \cite{Re,St}. An effective approach for this is building parametric stochastic models of the microstructure that provide realistic virtual samples whose physical properties can be computed numerically. Based on these results, it is then possible to study relationships between geometrical characteristics and descriptors of macroscopic physical properties, and, as a consequence, 
reduce the lab experiments needed to validate these relationships. On the other hand, when experimental datasets are available, one can generate further samples using the ideas of statistical reconstruction, see \cite{Il}.

In the present paper, we study 3D polycrystalline microstructures, which are interpreted as space-filling tessellations. There are various ways to model the latter \cite{CS,Ok}. We will focus on parametric 3D models of tessellations generated by stationary point processes. 
There are several types of such models. The basic Voronoi tessellation \cite{Ok} is often too simple to be used for fitting the grain boundaries of polycrystals \cite{QQ}. The more general Laguerre tessellation \cite{LZ}, on the other hand, became quite popular for the modeling of microstructures with approximately convex grains \cite{LY,SP}. More complex models exhibiting anisotropy or curved boundaries \cite{AL,SE} rely on higher-dimensional marks and are thus more difficult to handle.

When starting with the Poisson point process we obtain the Poisson-Voronoi or the Poisson-Laguerre tessellation \cite{La,Ok}, depending on whether we use a Voronoi or a Laguerre tessellation. 
More interesting random tessellations are obtained by replacing the Poisson point process with a Gibbs point process \cite{CS}. The Ord process \cite{OR} is one of the first references to this type of model. The crucial idea is to choose parametric potentials that allow us to control the geometrical properties of the tessellation in a prescribed manner. In \cite{DL} Gibbs-type Delaunay-Voronoi tessellations were investigated in 2D. In particular, methods of parameter estimation and goodness-of-fit testing have been suggested in \cite{DL} and applied to simulated data. 

Our aim is to extend the ideas of \cite{DL} in several ways in order to obtain new stationary models for grain microstructures of polycrystalline materials. Since the models are analytically intractable, our main tool of analysis will be stochastic simulation based on Markov chain Monte Carlo (MCMC). Such simulations enable us to investigate the properties of the models, which are presented mostly by histograms or kernel density estimates of distributions of basic geometrical characteristics. 
When going from 2D to 3D the computational demands increase,
and advanced algorithmic tools of computational geometry must be employed. Apart from that, we extend the model presented in \cite{DL} in two directions. First, the Gibbs-Laguerre tessellation is formally introduced and discussed. 
Secondly, we investigate a broader class of potentials, which allows us to control certain characteristics of the tessellation geometry. 
The goal is to provide clues to practitioners that allow to build stochastic models matching desired properties of geometrical characteristics that are important for applications. Examples could include the edge lengths in open-cell foams or the numbers of faces and vertices of grains in polycrystalline materials.

The present paper is organized as follows: In Section \ref{s1} we give some theoretical background of Gibbs point processes and Laguerre tessellations. Then various potential functions are defined and the properties of the corresponding Gibbs-Laguerre tessellations are investigated. 
In Section \ref{ssr} we present a simulation algorithm for Gibbs-Laguerre tessellations and explain our idea of statistical reconstruction.  This is followed in Section \ref{ss} by a simulation study with numerical results concerning the tessellation properties. In Section \ref{rd} the results of the previous sections are applied to the polycrystalline grain structure of an aluminum alloy sample obtained by synchrotron X-ray tomography \cite{PO}. First, we discuss the parameter estimation using standard methods for Gibbs processes. Then, as the main result of this paper, the statistical reconstruction of the microstructure of the sample based on a Gibbs-Laguerre model is presented. A discussion of the simulation outcome and concluding remarks are presented at the end of Section \ref{con}. Several complements and extensions to the paper are provided in online supplementary material, as mentioned throughout the text.

\section{Point processes and tessellations}
\label{s1}
\subsection{Gibbs point processes}
The microstructure of the materials that we intend to investigate is globally homogeneous. We are therefore interested only in stationary Gibbs point processes. The existence of such processes will be 
discussed in a separate paper using methods of \cite{DDG}. However, we observe the microstructure in a bounded window and, when taking an appropriate boundary condition, the finite volume Gibbs point process, cf. \cite{D19}, can be seen as an approximation of a stationary process. Therefore, in this paper we deal with finite point processes in the sense of \cite{MW}. Let $d > 1$ be an arbitrary fixed integer and $\Lambda\subset\R^d$ a bounded convex sampling window with $|\Lambda |>0,$ where $|\cdot |$ is the $d$-dimensional Lebesgue measure. Let $\pi^z_{\Lambda}$ denote the distribution of the restriction of a homogeneous Poisson point process on $\Lambda$ with intensity measure $\nu$ given by
\begin{equation}\label{intensity}
    \nu(B) = z |B|, \text{ for each Borel set } B \subset \R^d,
\end{equation}
where $z>0$ is called the intensity, cf. \cite{CS} for further details. If $z=1$, then we will write $\pi_{\Lambda} = \pi^1_{\Lambda}$ for short. Let $\bN$ be the family of all locally finite point configurations in $\R^d$ and ${\cal{N}}=\sigma(\{\x \in \bN: \card(\x \cap B)=m\}$, $B \subset \R^d$ Borel, $m\in \N \cup \{0\})$ the appropriate $\sigma$-algebra. Here, $\card(\x)$ denotes the number of points in $\x$. Further, $\bN^f$ is the family of all finite point configurations in $\R^d$ equipped with the trace $\sigma$-algebra ${\cal{N}}^f$ of ${\cal{N}}$, i.e., ${\cal{N}}^f = \{ \bN^f \cap N: N \in {\cal{N}} \}$, and $\bN^{f,k} \subset \bN^f$ is the family of configurations with exactly $k$ points. Finally, $\bN_{\Lambda}$ is the family of all finite point configurations in $\Lambda$. An energy function is a measurable function $E:\bN^f\longrightarrow\R \cup\{+\infty\}$, and we assume that it is nondegenerate, i.e., $E(\emptyset )<+\infty$. The energy function $E$ is said to be stable if there exists a constant $A \in \R$ such that 
\begin{equation} \label{stable}
E(\x)\geq A\cdot \card(\x) \text{ for every } \x \in \bN^f.
\end{equation}
 A finite Gibbs point process on $\Lambda$ with activity $z$ and energy function $E$ is a finite point process $\Phi$ having a density with respect to $\pi_{\Lambda}$ of the form
\begin{equation}\label{fdens}
    f( \x) = \frac{1}{Z_{\Lambda}} z^{\card(\x)} \exp{(-E(\x))}\quad\text{for } \x \in \textbf{N}_{\Lambda},
\end{equation}
where
\[ Z_{\Lambda} = \int_{\bN_\Lambda} z^{\card(\x)} \exp{(- E(\x))} \pi_{\Lambda} (d\x) \]
is a normalizing constant.

Analogously we define the notion of a marked Gibbs point process. Consider marks from a finite interval $I=(0,R_0]$ for some fixed $R_0>0$, a reference probability distribution $\gamma $ on $I$ and the Poisson process with distribution $\pi^z_{\Lambda \times I}$ on $\Lambda\times I$ and intensity measure $\nu\otimes\gamma$, where $\nu$ is given in \eqref{intensity}. 
The sets $\bN_I$, $\bN^f_I$, $\bN^{f,k}_I$ are defined analogously.
Then a finite marked Gibbs point process is given by the density \eqref{fdens}, but with respect to $\pi_{\Lambda \times I}$, where $\x\in\bN_{\Lambda \times I}$ are finite configurations of points in $\Lambda \times I$, and analogously in the formula for $Z_\Lambda$ one integrates over $\bN_{\Lambda \times I}$.

\subsection{Voronoi and Laguerre tessellations}
A tessellation in $\R^d$ is a locally finite system of space-filling closed sets, called cells, which are nonempty and have mutually disjoint interiors. 
We consider the tessellation $\mathfrak{T}(\x)$ generated by $\x = \{x_1,x_2,\ldots\} \in \bN$ or by $\x = \{x_1,x_2,\ldots\} \in \bN_I$. 
The cell $C_i$ corresponding
to the generator $x_i$,$i \in \N$, is defined with respect to some distance $\rho : \mathbb D \rightarrow \R_+$, where $\mathbb D=\R^d \times \R^d$ if $\x \in \bN$ or $\mathbb D=\R^d \times (\R^d \times I)$ if $\x \in \bN_I$, as
\begin{equation}\label{ce}
C_i = \{ x \in \R^d : \rho(x,x_i) \leq \rho(x,x_j) \text{ for all } x_j, i \neq j\}.
\end{equation}
\textit{Nonempty cells:} The definition of a tessellation allows only nonempty cells. Therefore, we assume that all generators of $\x$ create nonempty cells. Otherwise, we consider only the subset of $\x$ that generates nonempty cells, i.e., generators creating empty cells are excluded.

In particular, the choice of $\rho(y,x) = \lVert x-y \rVert$, $x,y \in \R^d$, where $\lVert \cdot \rVert$ denotes the Euclidean norm, results in the so-called Voronoi tessellation $V(\x)$, cf. \cite{Ok}. In this case no empty cells can arise.

\begin{figure}
    \center
    \includegraphics[width=5cm]{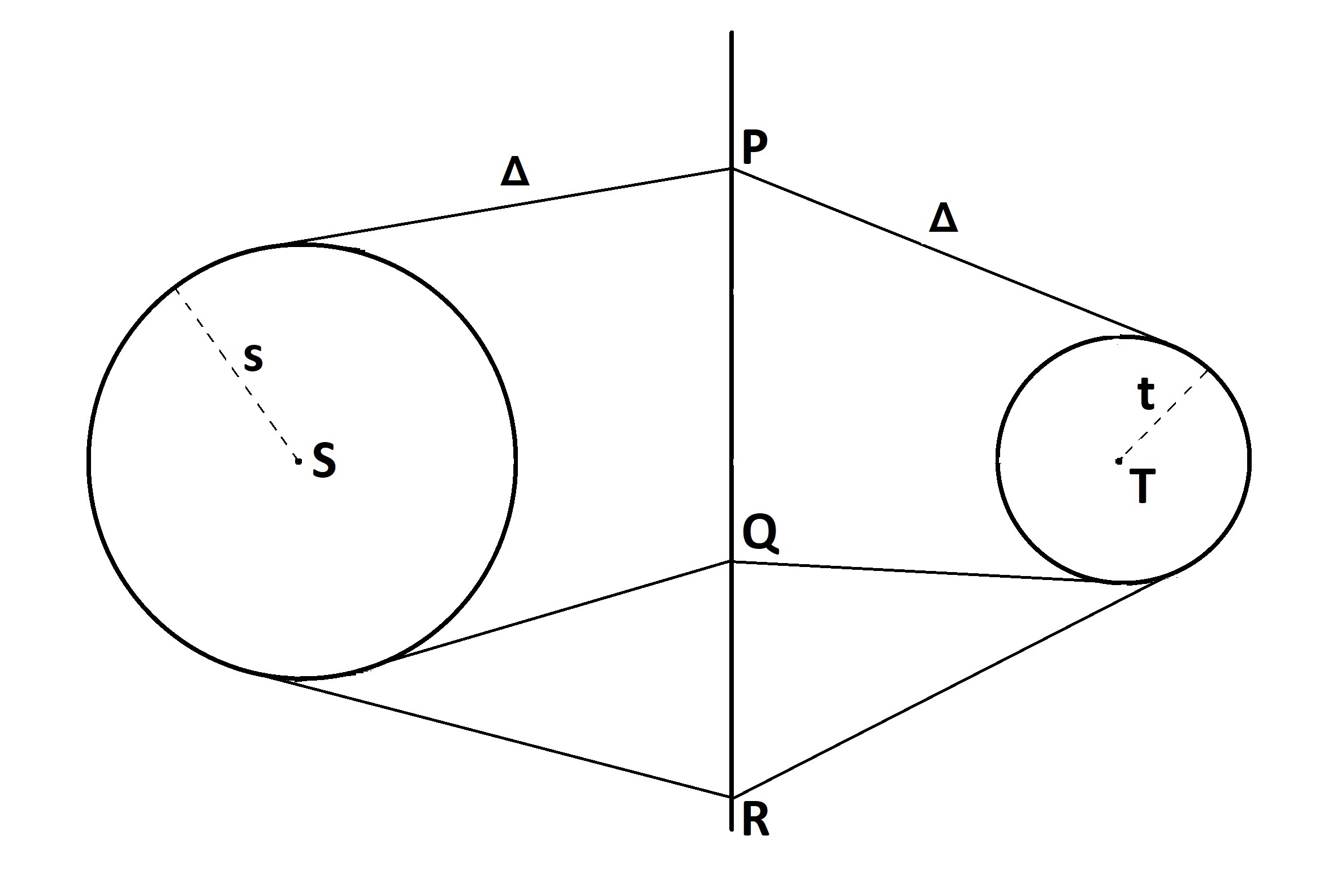} \\
    \caption{Illustration of the power distance given in \eqref{pow}. All three points P, Q, R have the same power distance with respect to circles with centers S, T and radii s, t, respectively. $\Delta $ is the square root of the power distance}
    \label{powerd}
\end{figure}

Another tessellation model can be defined for generators with marks. For $\x\in\bN_I$ and $(x,r)\in\x$, the power distance of a point $y \in \R^d$ with respect to the sphere $B(x,r)$ with center $x$ and radius $r$ is given by
\begin{equation}
    \label{pow}
    \rho(y,B(x,r)) = \lVert y-x \rVert^2 - r^2.
\end{equation}
The interpretation of the power distance is as follows: for each $y \in \R^d$ outside the sphere $B(x,r)$, the value $\rho(y,B(x,r))$ equals the squared length of the tangent line segment from $y$ to the sphere, cf. Fig.~\ref{powerd}. The distance $\rho(y,B(x,r))$ equals $0$ if $y$ is on the boundary of the sphere, and it is smaller than $0$ if $y$ is inside the sphere. The Laguerre tessellation $L(\x)$ is defined by choosing the power distance in the formula \eqref{ce}, cf. e.g., \cite{La}. 
 If a further generator $(x^{\prime},q) \in \x$ overlaps with $(x,r)$ over the center (i.e., $x^{\prime} \in B(x,r)$), it can happen that either the cell corresponding to marked point $(x^{\prime},q)$ does not cover $x^{\prime}$ or even that there is no cell at all (in this case the generator is omitted). Note that the Laguerre tessellations are invariant under transformations of radii of the form $r \mapsto \sqrt{r^2+t}$, where $t \in \R$ is fixed such that all radii remain positive. In the case that all radii are equal, the Laguerre tessellation reduces to the Voronoi tessellation. The cells of both Laguerre and Voronoi tessellations are convex polytopes, we denote $\mathcal{C}_d$ system of all convex polygons in $\R^d$.

\subsection{Random Gibbs-type tessellations.}\label{sec:randomGibbs}
In this paper we focus primarily on random tessellations in 3D Euclidean space, i.e., we set $d=3$ in the following, briefly $\mathcal{C}_3=\mathcal{C}.$ The system of generators forms a random point process. For stationary Poisson point processes with Euclidean distance, we obtain Poisson-Voronoi tessellations, for which closed analytical formulas are available for the moments of geometrical characteristics of cells, such as volume, number of faces, surface area, etc., cf. \cite{Ok}. For stationary Poisson point processes and the power distance given in \eqref{pow}, we obtain Poisson-Laguerre tessellations considered, e.g., in \cite{La}.
The aim of the present paper is to investigate Gibbs-Laguerre tessellations. The marked point process of generators will be a finite marked Gibbs point process in a bounded convex set $\Lambda\subset\R^3$ with marks coming from the interval $I$. 
Using periodic boundary conditions we obtain an approximation of a stationary model.
Note that Gibbs-Voronoi tessellations in $\R^2$ were studied in \cite{DL}. In a 3D tessellation we deal with $m$-dimensional facets, $m=0,1,2,3$ - namely, with vertices, edges, faces of cells, and the cells themselves. The energy function $E$ given in \eqref{fdens} is built as a sum of potentials (potential functions), see \cite{BA}, Def. 4.2.
A potential function $V$ is a measurable symmetric function,  $V: \bN^f_I \rightarrow \R \cup \{\infty\}$. In particular, $V_k: \bN^{f,k}_I \rightarrow \R \cup \{\infty\}$ is a potential function of order $k$, where
$k=1,2, \ldots, n$, $n = \card(\x)$, and $\x$ is a finite configuration of generators in $\Lambda \times I$. 
 We distinguish two types of potential functions. We speak about soft-core potentials if they are finite. On the other hand, hard-core potentials take on only one of the values $0$ or $+\infty$. In the rest of paper, when writing arguments of a potential function, we identify cells of tessellation with their generators.

\subsection{Periodic configuration}
Because of the bounded sampling window, edge effects have to be corrected, which requires knowledge of the process outside the window. 
This can be circumvented by employing periodic boundary conditions.  Without loss of generality, we assume
\[\Lambda=[0,1]^3,\,\x\in \mathbf{N}_{\Lambda \times I},\,\tilde{\x}=\cup_{(x,r) \in \x}\cup_{i\in {\mathbb Z}^3} (\tau_i(x),r),\]
where $\tilde{\x}$ is the periodic configuration on $\R^3 \times I$ and $\tau_i:\R^3 \rightarrow \R^3$ denotes the shift by $i$, $i \in \mathbb{Z}^3$. In the periodic setup a potential function of $k$-th order is summed over $k$-tuples of neighboring cells
in the periodic domain such that each periodic $k$-tuple makes a unique contribution. In other words there is only one contribution to the potential from all periodic images $\{(\tau_i(x),r): x \in C_1 \cup \ldots \cup C_k\}$, $i\in {\mathbb Z}^3$, of every $k$-tuple of neighboring cells $C_1,\ldots,C_k$.
For instance, for $k=1,2,3$ this is satisfied if the barycenter of the union of the $k$-tuple of cells belongs to the bounded set $\Lambda$. 
In general, a periodic energy function $\tilde{E} : \bN^f_I \rightarrow \R \cup \{\infty\}$ can combine different potentials of several orders and can be written in the parametric form (case of different orders)
\begin{equation} \label{parametric}
\begin{split}
\tilde{E}(\x) & =  V_{hard} +  \theta_1 \sum_{\stackrel{C\in {\mathfrak{T}}(\tilde{\x})}{\bary(C)\in \Lambda}}V_1(C) +  \theta_2 \sum_{\stackrel{C_1, C_2\in {\mathfrak{T}}(\tilde{\x}); C_1,C_2\sim_2}{\bary (C_1 \cup C_2)\in \Lambda}}V_2(C_1, C_2) + \ldots \\ &+  \theta_{n-1} \sum_{\stackrel{C_1, \ldots, C_{n-1} \in {\mathfrak{T}}(\tilde{\x})} { \stackrel{C_1,\ldots,C_{n-1}\sim_{n-1}} {\text{unique contribution}}}} V_{n-1}(C_1,C_2,\ldots,C_{n-1}) + \theta_n
V_n(C_1,C_2,\ldots,C_n), \\
\end{split}
\end{equation}
where all hard-core potentials are included in $V_{hard}$, $\theta_1, \ldots, \theta_n$ are real-valued parameters, $\bary(\cdot)$ denotes the barycenter of a given set and $\sim_{k}$ is the $k$-neighborhood relation. For $k=2,3$, neighboring cells are those which share a common face or edge, respectively.  For $k=n$, the entire tessellation is considered to be neighboring. It is important that each set of cells makes a unique contribution to the energy function. Note that there can be several potentials of the same order. Furthermore, the potential $V_{hard}$ can be written as a sum of hard-core potentials, i.e., potentials that can either be equal to zero or $+\infty$, i.e.,
\begin{equation*}
V_{hard} =  \sum_{\stackrel{C\in {\cal{T}}(\tilde{\x})}{\bary(C)\in \Lambda}}V_{1,hard}(C) + \ldots +
V_{n,hard}(C_1,C_2,\ldots,C_n).
\end{equation*}
Both energy function and periodic energy function are defined for $\x\in\bN^f_I$ only. If $E(\x)<+\infty $ or $\tilde{E}(\x)<+\infty$ in the periodic setup, we say that the configuration $\x $ is admissible.

\subsection{Examples of potential functions}\label{potfc}
We will deal with the following choices of potential functions. First, we consider the hard-core potential of first order:
\begin{equation}\label{enV1}
V_{1,hard}(C)= \begin{cases}
+\infty  & \text{if } h_{min}(C)\leq \alpha,\\
+\infty  & \text{if } h_{max}(C)\geq \beta, \\
+\infty  & \text{if } h_{max}^3(C) \geq B |C|, \\
0  & \text{else},
\end{cases}
\end{equation}
where $h_{min}(C),\,h_{max}(C)$ denotes the minimum, maximum distance between the cell barycenter and a face of $C,$ respectively, with $0<\alpha<\beta$, $B > 0.$ The parameter $\alpha$ forces the cells to be not too small, while $\beta$ forces them to be not too large. The parameter $B$ controls the shape of the cells---the smaller the value of $B$, the more regular are the shapes of the cells.

A soft-core potential of $k$-th order $V_k(C_1,\ldots,C_k)$ is a symmetric function of a $k$-tuple of neighboring cells. In practice, these potentials are often assumed to be nonnegative and bounded. 
In case there is no upper bound, an artificial bound $K > 0$ can be used ($K$ is some large constant depending on the particular potential). In 2D these two properties ensure the stability property \eqref{stable} of the energy function, cf. \cite{DDG} (unfortunately, this implication does not seem to be generally preserved in higher dimensions). 
A pair potential function studied later on is given by
\begin{equation}\label{enV2}
    V_{2,\VNR}(C_1,C_2)= \VNR(C_1, C_2) \wedge K
\end{equation}
with the neighbor-volume ratio (NVR) 
\begin{equation} \label{pairpo}
    \VNR(C_1, C_2) = \left(\frac{\max{\{|C_1|,|C_2|\}}}{\min{\{|C_1|,|C_2|\}}}-1\right)^{1/2}.
\end{equation}
When the potential given in \eqref{enV2} is multiplied by a real parameter $\theta$, the sign of $\theta$ is crucial. In the case when $\theta >0$, the neighboring cells tend to have a similar volume; on the other hand $\theta < 0$ forces the neighboring cells to have substantially different volumes.

In the following let $s:\mathcal{C}^k\rightarrow\R^k$ yield a vector of values of a geometrical characteristic assigned to a collection of cells $C_1,\dots C_k,$ in particular for $k=1,\,C\in\mathcal{C},$
$s(C) = \nof$ and $s(C) = \vol$ stand for the number of faces per cell and cell volume, respectively. In $s$ we do not distinguish the dimension of the domain, therefore $s(C_1,\dots ,C_k)=(s(C_1),\dots ,s(C_k))$. 
Let $T:\R^k\rightarrow\R $ be a functional of the sample $s(C_1,\ldots,C_k)$, namely $T(s(C_1,\ldots,C_k)) = \bar{s}(C_1,\ldots,C_k)$ stands for the sample mean computed over all cells in $\Lambda$, and, similarly, $s^2$ instead of $\bar{s}$ means the sample variance; 
$s_0\in\R$ is the value we want $T(s(\cdot))$ to take.

\begin{figure}
    \center
    \includegraphics[width=8cm]{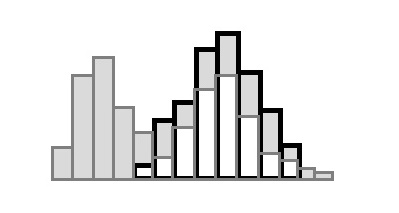}
    \caption{For the two histograms denoted in this figure by bars outlined in dark gray and black, their discrepancy is a constant based on a sum of contributions of the regions shaded in light gray}
    \label{fig:discrepancy}
\end{figure}

The potential of $n$-th order has a very special meaning. Recall that $n$ is the cardinality of the observed marked point pattern $\x$ (i.e., total number of cells) on the bounded sampling window $\Lambda \times I$. During simulations/reconstructions carried out below, this marked point pattern on $\Lambda \times I$ will change its cardinality; thus, $n$ is not constant in time. 
An example of a potential function of $n$-th order is
\begin{equation}\label{enVN}
V_{n,T}^s(C_1,\ldots,C_n)=\left(|T(s(C_1,\ldots,C_n)) - s_0|\right)^{1/2}.
\end{equation}
Later on, potential functions of the form \eqref{enVN} will be referred to as reconstructing potentials.
A special case of this potential of $n$-th order (with $T(s(C_1,\ldots,C_n))=\dsc(H_{s(C_1,\ldots,C_n)},H_s^{\prime})$ and $s_0=0$, where $\dsc$ is an abbreviation of discrepancy, which is defined below, see \eqref{dsc}) allows us to control not only the moments of some geometrical characteristic but also its entire distribution. More precisely, it is given by
\begin{equation}\label{enVN2}
V_{n,\dsc}^s(C_1,\ldots,C_n)=\left(\dsc(H_{s(C_1,\ldots,C_n)},H_s^{\prime})\right)^{1/2},
\end{equation}
where $H_{s(C_1,\ldots,C_n)}$ is the histogram of the chosen geometrical characteristic computed from all cells, and $H_s^{\prime}$ is the prescribed targeting histogram of $s$ that we want to approach (this can be typically obtained from data).
Alternatively, we could deal with the cumulative histogram or empirical distribution function instead of the histogram.
Let us consider some geometrical characteristic of the $m$-dimensional facets, $m=0,1,2,3$, taking values in an interval $[a,b]$. For some integer $J$, let $D = \{t_i\}_{i=0}^J$ with $t_i < t_{i+1}$ for all $i$ be a decomposition of the interval $[a,b]$ into $J$ subintervals such that $t_0=a$ and $t_J=b$ ($D$ does not need to be equidistant). Each histogram $H$ can then be represented by some numbers $h_1, \ldots, h_J$  interpreted as frequencies of the classes $1, \ldots, J$ (i.e., $h_i$ is the number of facets for which the value of the considered geometrical characteristic belongs to the interval $[t_{i-1},t_i)$). Using the abbreviating notation $S = \sum_{i=1}^J h_i$, the discrepancy between a pair of histograms $(H, H^{\prime})$ defined over the same interval and having the same bins (this implies the same number of classes) can be written as
\begin{equation} \label{dsc}
\dsc(H, H^{\prime}) = \sum_{i=1}^J \left|\frac{h_i}{S} - \frac{h^{\prime}_i}{S^{\prime}}\right|.
\end{equation}
An illustration of $\dsc$ can be seen in Fig.~\ref{fig:discrepancy}. The discrepancy $\dsc$ measures the difference between two histograms and is minimized when they are identical up to some positive multiplicative constant (i.e., there exists a constant $M>0$ such that $h_i = M h^{\prime}_i$ for every $i=1, \ldots, J$). If we omit the normalizations $S, S^{\prime}$ in the definition of discrepancy, given in \eqref{dsc}, then the discrepancy is minimized if the two histograms are identical.

\section{Simulation and statistical reconstruction}
\label{ssr}
This section deals with two closely related topics, simulation and statistical reconstruction of marked point processes of tessellation generators on a bounded and convex observation window $\Lambda \times I$. 
We will use the
Markov chain Monte Carlo (MCMC)-- namely, the Metropolis-Hastings birth-death-move algorithm (MHBDM), cf. \cite{DL,MW}. An adaptation of the evolution step of this algorithm for marked Gibbs point processes is presented in Section \ref{sec:BDMA}. It is the key step of the algorithm and we treat it separately, since we will use it in both simulation and reconstruction tasks. The MHBDM algorithm for simulation of the Gibbs-Laguerre tessellation itself is presented in Section \ref{sec:simulationGibbsLaguerre}. Its goal is to generate a stationary Markov chain of marked point configurations that tends to the target distribution.

On the other hand, the statistical reconstruction is based on a given data pattern, and the aim is to generate samples that are statistically similar \cite{Il}. 
In Section \ref{sec:reco} we propose an alternative approach to statistical reconstruction using the evolution step of MHBDM, in which an auxiliary parameter is used. 
This method is compared in Section \ref{reco_comp} with a classical reconstruction based on greedy algorithm. This comparison is postponed since it requires an experimental data.
Finally, Section \ref{tech} discusses the importance of local recomputations of the tessellation, which are necessary to achieve a substantial reduction in computing time.

\subsection{Evolution step of the birth-death-move algorithm}\label{sec:BDMA}

The general form of the MCMC Metropolis-Hastings birth-death-move algorithm is described in \cite{MW}. Recall that admissibility of a finite point configuration $\x$ means that its energy  $E(\x)$ is finite. Let $f$ be defined as in equation \eqref{fdens}.
The evolution step of the MHBDM algorithm for an admissible $\x_0 \in \Lambda \times I$ with $n = \card(\x_0)$ can be written as follows.
\paragraph{Algorithm 1} (evolution step of MHBDM). \\

Do one of the following (with probability $\frac{1}{3}$ each):
        \begin{enumerate}
            \item[(a)] ``birth": generate a point $y$ uniformly in $\Lambda$ (i.e., $y \sim U(\Lambda)$) and a radius $r \sim U(I)$ and set
                \[ \x_1= \begin{cases}
                    \x_0\cup\{(y,r)\} & \text{with probability } \min\left(1,\frac{f(\x_0\cup\{(y,r)\})}{(n+1)f(\x_0)}\right), \\
                    \x_0 & \text{otherwise;}
                \end{cases} \]
            \item[(b)] ``death": choose a point $(y,r)$ from $\x_0$ at random and set
                \[ \x_1= \begin{cases}
                    \x_0\setminus\{(y,r)\} & \text{with probability } \min\left(1,\frac{nf(\x_0\setminus\{(y,r)\})}{f(\x_0)}\right), \\
                    \x_0 & \text{otherwise;}
                \end{cases} \]
            \item[(c)]\label{item:MHBDMmove} ``move": choose a point $(y,r)$ from $\x_0$ at random and generate $x \sim N_3(y,\Sigma)$ with the covariance matrix $\Sigma$, $s \sim U(I)$ and set
                \[ \x_1= \begin{cases}
                    (\x_0\setminus\{(y,r)\})\cup \{(x,s)\} & \parbox[t]{.4\textwidth}{$\text{with probability}$ $\min\left(1,\frac{f((\x_0\setminus\{(y,r)\})\cup \{(x,s)\})}{f(\x_0)}\right)$,} \\
                    \x_0 & \text{otherwise.}
                \end{cases} \]
        \end{enumerate}
Here, 
\begin{equation}\label{rsig}
\Sigma=diag\{\sigma^2,\sigma^2,\sigma^2\},\,\sigma >0,\end{equation} 
$U$ denotes the uniform distribution and $N_3$ denotes the trivariate Gaussian distribution. All proposals are sampled independently of each other.
Using the Gaussian distribution for the move proposal distribution is a common choice, cf. \cite{DL}.
The constant $\sigma $ of the proposal distribution will be chosen appropriately.
Throughout the rest of this paper the observation window $\Lambda = [0,1]^3$ is used.
First note that in step (c) of the algorithm, the point $x$ can always be considered to belong to $\Lambda$; if $x \notin \Lambda$ then the periodic image of $x$ in $\Lambda$ is taken. 
The second observation is that with formula \eqref{fdens} the acceptance ratios $H_k$, $k=$ ``birth", ``death", ``move", in the steps (a),(b),(c), are of the form
\begin{equation} \label{hastings} H_{k} = c_{k} \cdot \exp(E_{b}-E_{a}),
\end{equation}
where $E_{a}$ is the energy of the proposal (e.g., $\x_0\cup\{(y,r)\}$ in step (a)), $E_{b}$ is the energy of $\x_0$ and the constant $c_{k}$ is equal to $\frac{z}{n+1}$ in step (a), $\frac{n}{z}$ in step (b) and $1$ in step (c).
Note that during the ``death" step it is possible to exclude one generator, but an even higher number of generators can be deleted in the ``birth" or ``move" steps. The reason for this is that during the latter two steps empty cells can arise
whose generators are thus removed.
\\ \\
\textit{Choice of constants:} In all simulations/reconstructions using the evolution step of MHBDM, \textit{Algorithm 1}, we set $R_0=0.2$ as the upper bound for the marks, $\sigma$ to $0.015$ in \eqref{rsig} and the activity $z$ to the fixed value $2000$.

\subsection{Simulation of Gibbs-Laguerre tessellations}\label{sec:simulationGibbsLaguerre}
An MHBDM algorithm simulating Gibbs-Voronoi tessellations in $\R^2$ is presented in \cite{DL}.
We extend their work to Gibbs-Laguerre tessellations in $\R^3$.
Pseudocode for our simulation algorithm of a point configuration of generators is given below.

\paragraph{Algorithm 2} (Simulation via MHBDM). 
\begin{enumerate}
    \item construct an admissible marked point configuration $\x_0$,
    \item $n \leftarrow \card(\x_0)$,
    \item run the \textit{Algorithm 1} (taking $\x_0$ and yielding $\x_1$),
    \item $\x_0 \leftarrow \x_1$,
    \item repeat steps 2 to 4 $(S-1)$ times,
    \item return $\x_0$.
\end{enumerate}

The marked point patterns obtained in $\Lambda \times I$ 
can be transformed into tessellations that are considered to be 
samples of Gibbs-Laguerre tessellations.
Various models based on different densities $f$ in \eqref{fdens} will be 
considered.
Clearly, the number of iterations $S$ depends on the considered model; more complex models tend to require more iterations in \textit{Algorithm 2} to approach the target distribution.
The convergence of the basic MHBDM algorithm was proven under mild conditions in \cite{MW}, Section 7.3. As claimed in \cite{DL}, Section 3.2, for models with hard-core potentials the convergence of the algorithm is difficult to prove in cases when the tessellation model becomes too rigid. This might happen if the NVR potential \eqref{pairpo} is multiplied by a large positive parameter $\theta$, which we try to avoid in the following.


\subsection{Reconstruction approach}\label{sec:reco}
The aim of the statistical reconstruction of point patterns as introduced in \cite{TS} (see \cite{Il} for a textbook version) is to generate point patterns with distributional characteristics close to those of a given point pattern (data). 
This is carried out by fixing the number of points in the observation window and by running some iterative optimization method (e.g., greedy algorithm, \cite{CL}, Chapter 16, or simulated annealing, \cite{Laarhoven1987}) that modifies a single point in each step to minimize the discrepancy to the data. The resulting point pattern is taken as a reconstruction of the data pattern.


In the present paper we develop an alternative approach, which is based on the Gibbs point process simulation discussed above. Here, the energy function is chosen in such a way that the states of the corresponding Markov chain in the stationary regime are marked point patterns having characteristics close to those of the given data.

For this purpose, we simplify the equation in \eqref{parametric} such that the periodic energy function is written as the sum of $k$ reconstructing potentials of $n$-th order, $V_{n,T_1}^{s_1},\ldots, V_{n,T_k}^{s_k}$, cf. \eqref{enVN}. The potentials differ from each other in the choice of geometrical characteristic $s_i$ and functional $T_i$, $i=1,\ldots,k$. A hardcore potential might be included as well, i.e., 
\begin{equation} \label{energy_r}
\begin{split}
\tilde{E}(\x) & =  V_{hard} + \sum_{i=1}^k \theta_n^i
V_{n,T_i}^{s_i}(C_1,C_2,\ldots,C_n). \\
\end{split}
\end{equation}
Besides $V_{hard}$, information regarding the point pattern of the data is contained also in the reconstructing potentials $V_{n,T_i}^{s_i}$--namely, in the constants $s_{0,i}$ and the histograms $H_{s_i}^{\prime}$ in the formulas \eqref{enVN} and \eqref{enVN2}, respectively. The energy function is parametrized by a vector $\boldsymbol{\theta} = (\theta^1_n,\ldots,\theta^k_n)$ of so-called control parameters.
If $\theta^i_n > 0$ for all $i \in \{1,\ldots,k\}$ then the entire energy function is nonnegative and its minimum is greater or equal to $0$.
The vector of control parameters is used for specifying the precision of the reconstruction, 
while the activity parameter $z$ can be fixed. 
The \textit{Algorithm 1} yields $\x_1$ with the tendency that $E(\x_1)$ is smaller than $E(\x_0)$, which is evident from the formula of acceptance ratios \eqref{hastings}. 
Despite this, the MHBDM algorithm cannot be used for the direct minimization of the energy function. 
How much we are able to decrease the energy depends on the parameter $\boldsymbol{\theta}$. This issue will be discussed in Section \ref{control} below.
Instead of a fixed number of iterations, we use a different stopping condition after which the algorithm is terminated. 
The algorithm ends if there is no significant change larger than some $\delta > 0$ of the energy function during a series of $t$ steps. We call the pair $(\delta,t)$ the stopping criterion.
The reconstruction algorithm (as used in
Section \ref{rd}) is given as follows.



\paragraph{Algorithm 3} (Reconstruction via MHBDM). 
\begin{enumerate}
    \item construct an admissible marked point configuration $\x_0$, 
    \item $n \leftarrow \card(\x_0)$, 
    \item run the \textit{Algorithm 1} (taking $\x_0$ and yielding $\x_1$),
    \item $\x_0 \leftarrow \x_1$,
    \item if the energies of the last $t$ marked point configurations obtained by step 3 do not differ more than $\delta$, then return $\x_0$, else goto 2. 
\end{enumerate}

Note that stopping conditions other than the pair $(\delta,t)$ are possible, e.g., terminate the algorithm if the energy of the point pattern decreases below some threshold. This approach could lead to a substantial decrease in computational time without necessarily ensuring that the algorithm arrives in a stationary regime. Finally, note that a stopping criterion must be suggested carefully; otherwise, it may not terminate at all. A related numerical study is presented in the online supplementary material.
\subsubsection{Control parameters} \label{control}
The control parameters enable us to influence the accuracy of the reconstruction. For the ease of explanation assume $k=1$, i.e., $\boldsymbol{\theta} = \theta_n$, and omit the hardcore potential $V_{hard}$. For practical reasons consider $\theta_n > 0$ only (a negative value would increase the energy instead of decreasing it). Assume that there is a stationary distribution to which the algorithm converges. Then the algorithm generates a Markov chain whose states in the stationary regime have energies oscillating around some mean value $L \geq 0$, which is a measure of the accuracy of the reconstruction. More specifically, $L$ corresponds to the mean Euclidean distance for the moments of the characteristics in the case of the potential given in \eqref{enVN} or to the mean discrepancy measure for histograms in the case of \eqref{enVN2}. 
The accuracy improves, i.e., $L$ decreases, with increasing values of the control parameter $\theta_n$. Note that if $k>1$, the potentials are competing and the situation becomes more complicated.
The value of the control parameter must not be too high, however, since otherwise only very few changes are accepted during the run of the algorithm (the acceptance probabilities of non-improving suggestions tend to zero), and it is difficult for the Markov chain to reach the stationary regime. This is demonstrated by the following example. \\

\textit{Example:}\label{sec:potentialExample}
Consider an energy function with a single potential $V^s_{n,T}$ given in \eqref{enVN}, where $s$ is the number of cell faces ($\nof$), $T$ is the sample mean and $s_0=12$, i.e., the aim is to get tessellations with mean number of faces per cell equal to $12$. 
In the $i$-th step of the algorithm, let the sample mean of the number of cell faces be equal to $14.258$. 
Suppose that a move of a generator is suggested and the sample mean after the suggested operation is $14.264$. The choice of $\theta_n$ influences the probability of acceptance; more precisely, recall from step (c) of \textit{Algorithm 1} that the acceptance probability is given by $\frac{f((\x_0\setminus\{(y,r)\})\cup \{(x,s)\})}{f(\x_0)}$, where $\x_0$ is the point pattern in the $i$-th step and the point $(y,r) \in \x_0$ is suggested to be substituted by $(x,s)$.
The energy of this change is $\theta_n \cdot \left((14.258-12)^{1/2}-(14.264-12)^{1/2}\right) = - 0.002 \cdot \theta_n$, and the acceptance ratio is proportional to $e^{- \theta_n \cdot 0.002}$.
 The acceptance probability equals $e^{-0.2}=0.819$ or $e^{-2}=0.135$ for $\theta_n = 100$ or $\theta_n = 1\,000$, respectively. An increase to $\theta_n = 10\,000$ results in an almost vanishing acceptance probability of $e^{-20}= 2.06 \cdot 10^{-9}$. \\ 
\indent Because a better fit is achieved with increasing $\theta_n$, it seems to be undesirable to estimate the parameter $\theta_n$ by the methods described in the online supplementary material, and we simply advise taking the smallest value of the control parameter that yields a satisfactory accuracy of reconstruction. 
The value of $\theta_n$ influences the fluctuations in energy, as well. The fluctuations are smaller when the value of the parameter is higher, which causes greater penalization of non-improving suggestions. This must be kept in mind when applying the stopping criterion $(\delta,t)$. 
\subsubsection{Activity} 
The most natural option for the activity $z$ is to set it to the total number of points in the pattern that is being reconstructed. However, if the potential \eqref{enVN2} defined by the discrepancy of histograms of cell volumes is included in the energy function, the choice of the activity $z$ does not play a significant role. It can be fixed arbitrarily, 
as long as the value remains on the same order of magnitude as
the intensity of the reconstructed point pattern. The reason for this is that the cell volumes and the intensity of the Gibbs point process are strongly correlated. So, assuming that the control parameter $\theta_n$ is large enough, the influence of the reconstructing potential exceeds that of the activity $z$.

A detailed description of the model's behavior with respect to various choices of parameter values is presented in Section \ref{rd}, where the reconstruction of experimental data is discussed.

\subsection{Computational geometry aspects}\label{tech}
When dealing with the simulation of Gibbs-Laguerre tessellations or with the reconstruction of Laguerre tessellations, 
suitable implementations of efficient geometrical data structures and algorithms are needed.
We use the open source software Voro++ \cite{Ry} for the computation of Voronoi and Laguerre tessellations. 
Periodic boundary conditions are also handled by this library.


The geometries of the tessellations are needed in each call of \textit{Algorithm 1} in order to determine the values of the potentials and the energy function. Computing the entire tessellation in each step would be very time consuming and inefficient. 
Each proposed tessellation differs from the previous one only by a small number of cells.
These have to be recomputed in order to determine the change in the value of energy, while the rest is kept unchanged. 
An algorithm for finding the set of cells that needs to be updated is described in \cite{QU}. 
Recomputing the energy function only locally drastically reduces the runtime of each iteration of the algorithm, and it is one of the main reasons why the MCMC simulations approach the target distribution in a reasonable time.

\section{Numerical studies}\label{ss}
This section consists of two parts: an investigation of the neighbor-volume ratio in Section~\ref{sec:pairPotentialVNR}, which is an example of a (2nd-order) pair potential, and a comparison of Gibbs-Laguerre and Poisson-Laguerre tessellations in Section~\ref{sec:ComparisonPLT}. 
The first part provides a deeper insight into the usage of the pair potential. This potential is parametrized and treated on its own with hardcore conditions, or together with a reconstructing potential based on the histogram of the number of faces. 
The second part concerns the comparison of realizations of Gibbs-Laguerre tessellations obtained by \textit{Algorithm 2} and those drawn from Poisson-Laguerre models. It emphasizes the capability of the Gibbs approach to generate tessellations manifesting greater variability in cell shapes.

\textit{Note about notation:} Later in Sections \ref{ss} and \ref{rd}, we will need to describe briefly what a model, i.e., its energy function, looks like. This will be accomplished by listing incorporated potentials and parameters, collected in Section \ref{potfc}. For all potentials and soft-core parameters, the first subscript denotes the order of the potential. There can be more than one potential of the same order; therefore the corresponding soft-core parameters will be distinguished by superscripts. 
Table 1 gives an overview of all considered Gibbs-Laguerre tessellations.
For the reconstructing potentials \eqref{enVN} we will use the abbreviating notation $T(s)=\bar{s}$, $T(s)=s^2$ and $T(s)=\dsc(H_s,H^{\prime}_s)=\dsc$ to state in which statistic of the sample we are interested. The symbols $\nof$ and $\vol$ abbreviate the number of faces per cell and the cell volume, respectively. For example, the term
$$V_{n,\dsc}^{\nof}; s=\nof, T(s) = \dsc(H_s, H_s^{\prime}), H_{\nof}^{\prime}, \theta_n$$ 
denotes the model with a single reconstructing potential defined by the number of faces per cell as the geometrical characteristic $s$, histogram discrepancy as the functional $T(s)$, prescribed histogram $H_{\nof}^{\prime}$ and a soft-core parameter $\theta_n$.


\begin{table}
\center
\caption{Potentials and parameter specifications of the considered Gibbs-Laguerre tessellations and their identifying labels. The symbols $RTk$, $k=1,\ldots,7$, denote either a single random tessellation or a class of random tessellations sharing a common choice of potentials. The missing random tessellations can be found in the online supplementary material. The rightmost column gives number of parameter specifications used in the given tessellation, and ``data" means that real microstructure data, cf. Fig.~\ref{expdata}, are used in the model}
\label{Tno}
\begin{tabular}{ccccc}
\hline
\multirow{2}{*}{Tessellation(s)} & \multirow{2}{*}{Potential(s)} & \multirow{2}{*}{Label} & \multirow{2}{*}{Parameter(s)} & Number of \\
 &  & &  & specifications\\
\hline
\vspace{0.1cm}
$RT1$ & $V_{1,hard}+V_{2,\VNR}$ & \eqref{model3} & $\alpha, \beta; \theta_2$ & 2 \\
\vspace{0.1cm}
$RT2$ & $V_{1,hard}+V_{2,\VNR}+V_{n,dsc}^{\nof}$ & \eqref{model4} & $\alpha, \beta; \theta_2, \theta_n$ & 2 (data) \\
\vspace{0.1cm}
$RT6$ & $V_{n,dsc}^{\nof}$ & \eqref{model8} & $\theta_n$ & 6 (data) \\
\vspace{0.1cm}
$RT7$ & $V_{n,dsc}^{\nof}+V_{n,dsc}^{\vol}$ & \eqref{model10} & $\theta_n^1,\theta_n^2$ & 12 (data)\\
\hline
\end{tabular}
\end{table}

\subsection{Neighbor-volume ratio}\label{sec:pairPotentialVNR}
The pair potential $V_{2,\VNR}$, given in \eqref{enV2} and studied in 2D in \cite{DL}, introduces interactions between neighboring cells. The strength of interactions is influenced by the parameter $\theta_2$ given in \eqref{parametric}. When $\theta_2$ is positive, we will call the model regular. The potential is minimized when both cells in a pair of neighboring cells have the same volume, and consequently when all cells have the same volume. On the other hand, if $\theta_2$ is negative, then pairs with totally different volumes are preferred, and we will speak about an irregular model. In the irregular case, there are often many more cells than in the regular case, given a fixed value of $z.$ 
Whether $\theta_2$ is positive or negative has a strong effect on the number of cells in $\Lambda$.
Hardcore parameters, e.g., \eqref{enV1}, can be used to reduce $\Lambda$ significantly (in particular by using the bounds $\alpha$ and $\beta$). The class $RT1$, cf. Table \ref{Tno}, consists of two random tessellations given by
\begin{equation}  \label{model3}
\alpha = 0.02, \quad \beta = 0.095, \quad \theta_2=\pm 1.
\end{equation}

\begin{figure}
    \centering
    \includegraphics[width=5.5cm]{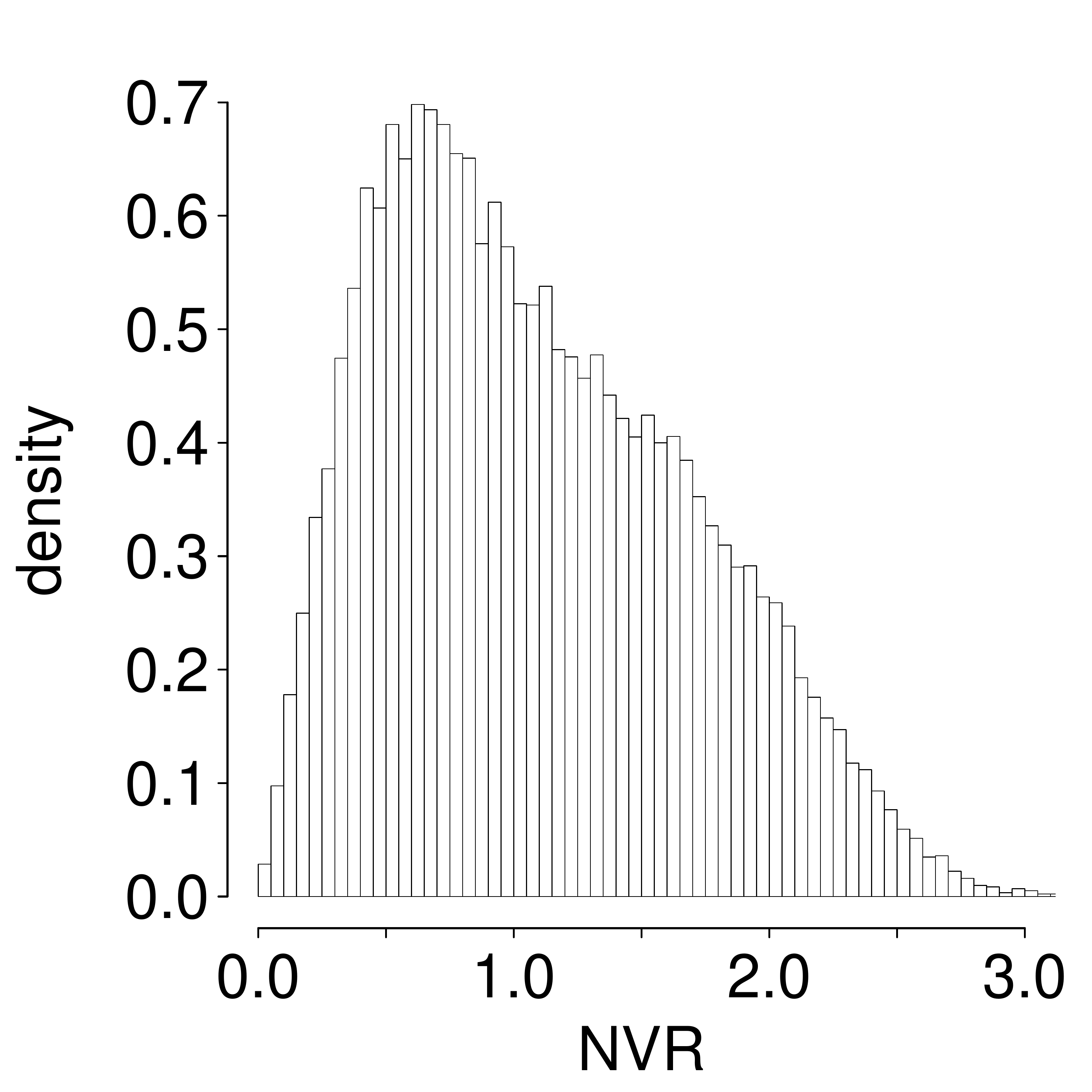}
    \includegraphics[width=5.5cm]{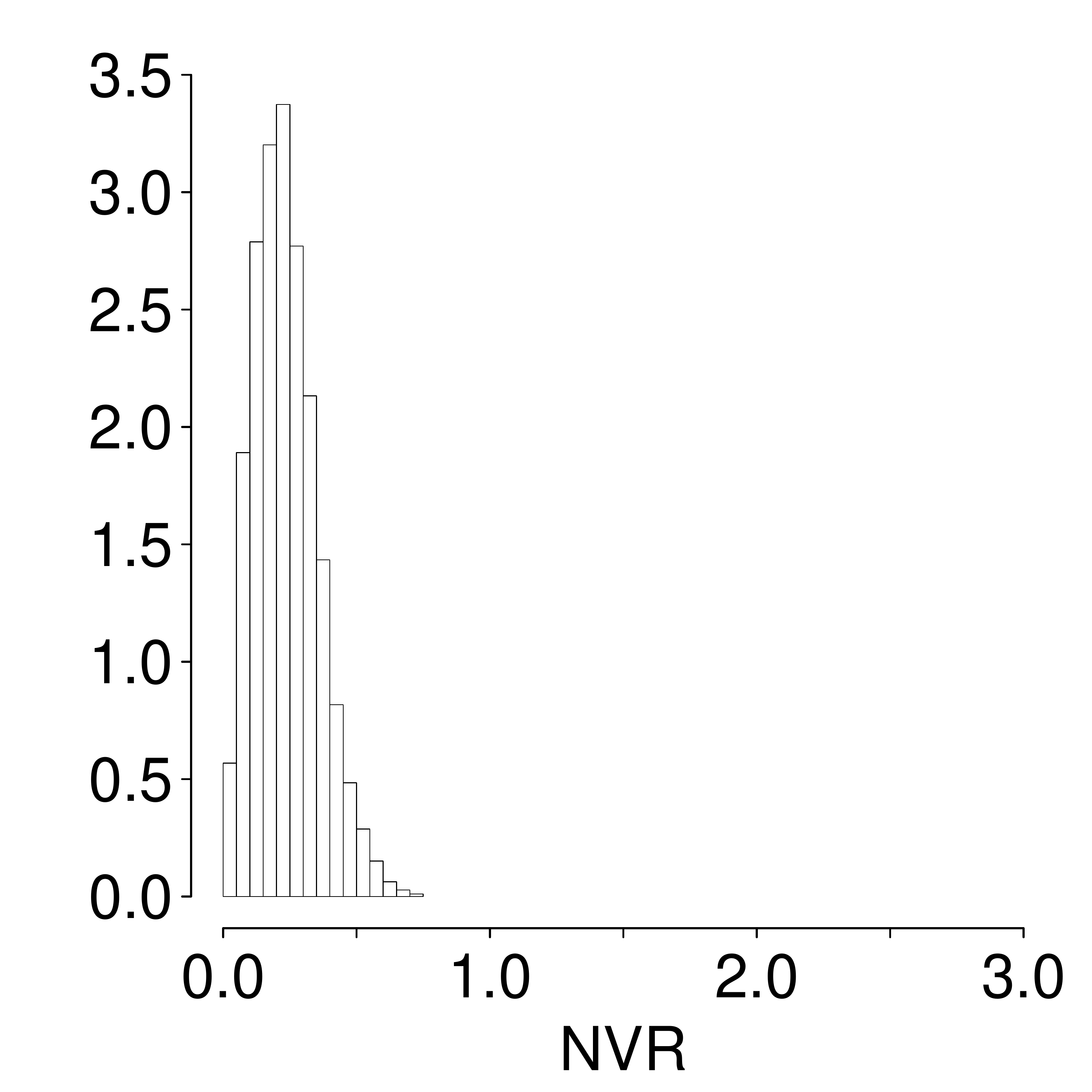}\\
    \text{a) \hspace{17em} b)}
    \caption{Histogram of relative frequencies of the NVR statistic for the simulated tessellation of class $RT1$: (a) irregular ($4497$ cells) and (b) regular ($752$ cells) specification}
    \label{plot3}
\end{figure}

Note that the pair potential given in \eqref{enV2} can be combined with other potentials, e.g., with the potential given in \eqref{enVN2}. The combination then shares properties of both components. The potential in \eqref{enVN2} is minimized if the histogram of the target characteristic of the tessellation approaches the prescribed histogram. This way we can control the distribution of the chosen characteristic. The parameter $\theta_n$ controls how closely the distribution of the target characteristic of the tessellation matches the prescribed histogram. As mentioned in Section~\ref{sec:potentialExample}, the value of $\theta_n$ must be reasonably high. Thus, the class $RT2$, cf. Table \ref{Tno}, consists of two random tessellations given by
\begin{equation}   \label{model4}
    \begin{array}{c}
        \alpha = 0.02, \quad \beta = 0.095, \\
        s=\nof, \ T(s)=\dsc(H_s,H_s^{\prime}), \\
        \theta_2=\pm 1,  \quad \theta_n=100\,000,
    \end{array}
\end{equation}
where $H^{\prime}_{\nof}$ is set to be the histogram of the number of faces of the experimental dataset described in Section~\ref{sec:experimentalData}, see Fig.~\ref{stats}a.

\begin{figure}
    \centering
    \begin{tikzpicture}
        \node (plots1) { 
        \includegraphics[width=0.4\linewidth]{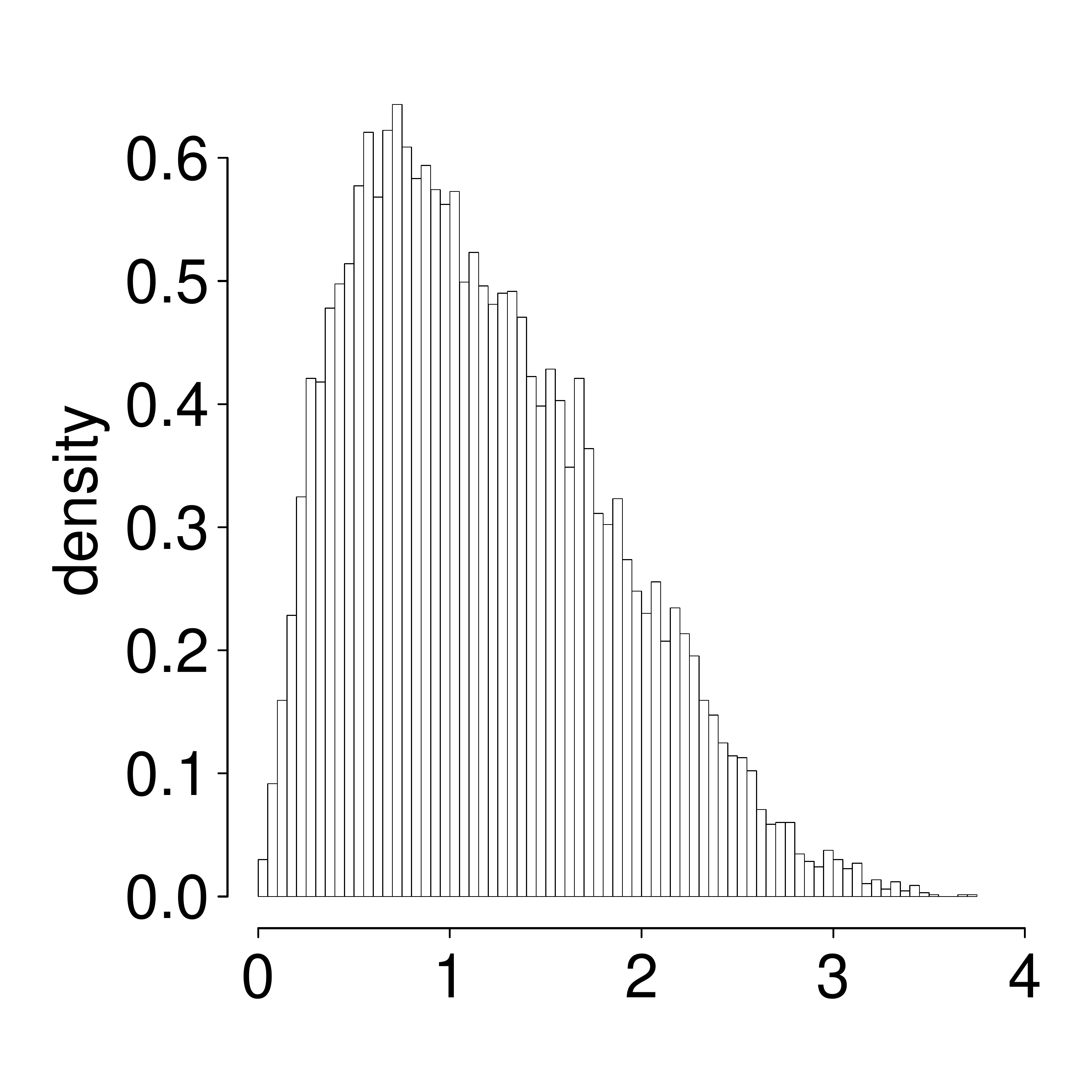}
        \includegraphics[width=0.4\linewidth]{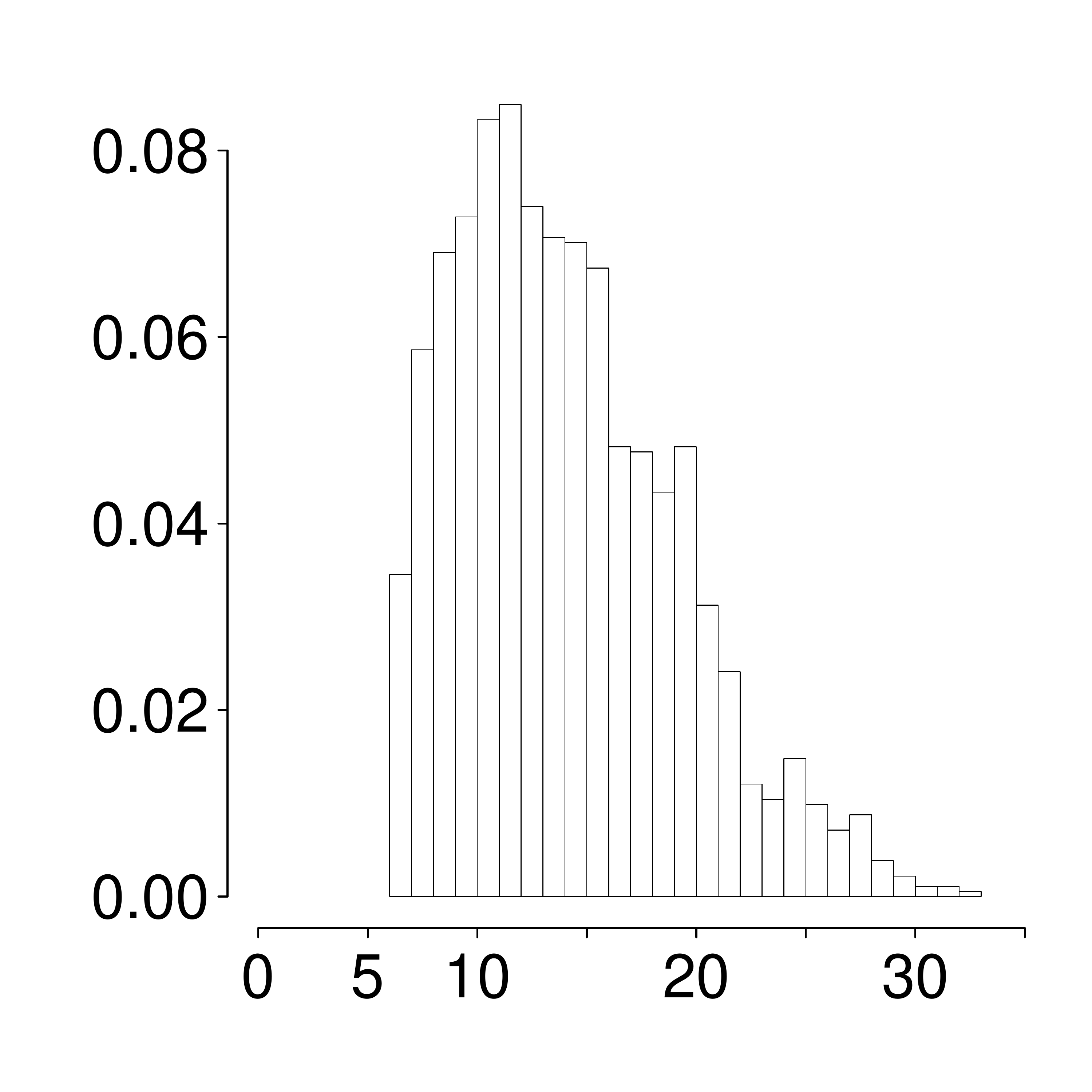} 
        }; 
        \node[anchor=base, yshift=-2ex] at ($(plots1.north west)!0.25!(plots1.north east)$) { A };
        \node[anchor=base, yshift=-2ex] at ($(plots1.north west)!0.75!(plots1.north east)$) { B };

        \node[xshift=-2em]at ($(plots1.north west)!0.5!(plots1.south west)$) { I };
    \end{tikzpicture}
    \begin{tikzpicture}
        \node (plots2) { 
        \includegraphics[width=0.4\linewidth]{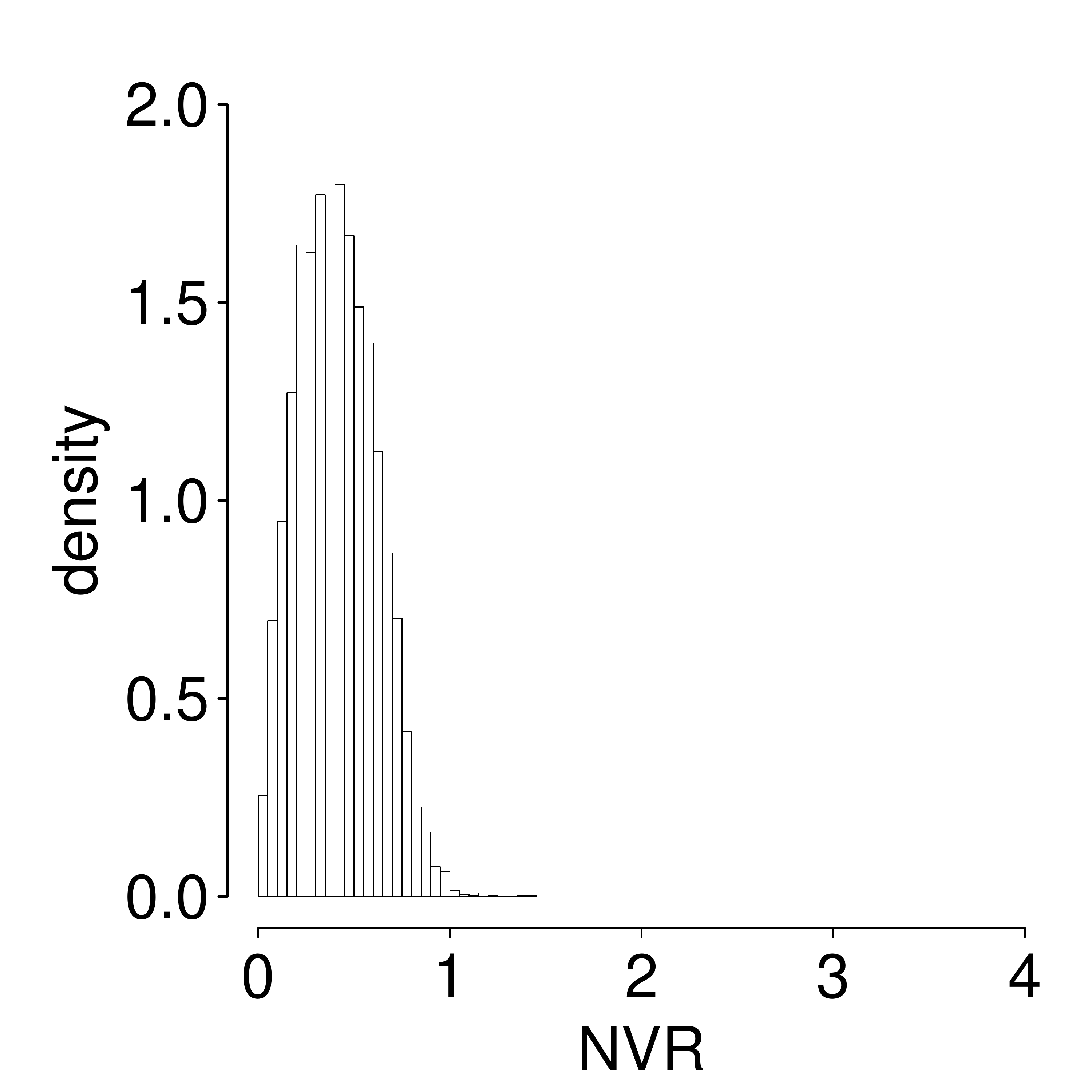}
        \includegraphics[width=0.4\linewidth]{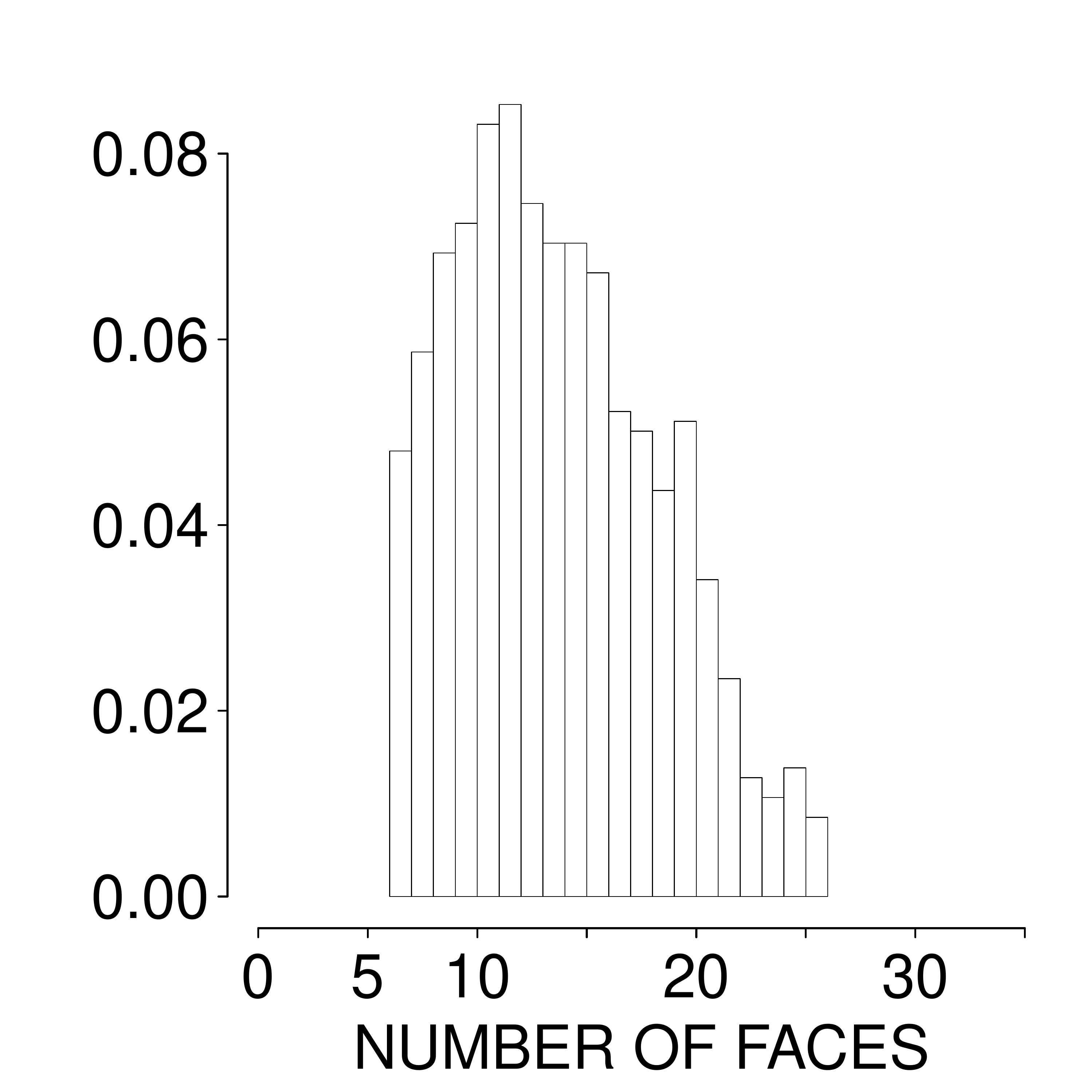}
        };
        \node[xshift=-2em] at ($(plots2.north west)!0.5!(plots2.south west)$) { II };
    \end{tikzpicture}
    \caption{Characteristics for two simulated tessellations from the class $RT2$: the histograms of relative frequencies of the NVR statistic are plotted in column A, the histograms of relative frequencies of the number of faces in column B; results for an irregular tessellation ($3166$ cells) are provided in row I and for a regular tessellation ($956$ cells) in row II}
    \label{plot4}
\end{figure}

In the tessellation generated by an irregular model the variance of the neighbor-volume ratio given in \eqref{pairpo} is much larger than in the regular case, as the neighboring cells tend to have significantly different volumes (cf. the histograms in Fig.~\ref{plot3}). The tessellations of the class $RT2$ demonstrate that the combination of an interaction potential and a reconstructing potential can work successfully: the properties observed in the case of $RT1$ are preserved, and the discrepancy between each of the histograms in column B of Fig.~\ref{plot4} and the corresponding histogram coming from experimental data, Fig. \ref{stats}a, is small. Note that realizations of all random tessellations in $RT1$ and $RT2$ were obtained after three million steps of \textit{Algorithm 2}.

\subsection{Comparison with Poisson-Laguerre tessellations (PLT)}\label{sec:ComparisonPLT}
In contrast with Poisson type tessellations, the Gibbs point process allows for ready modification of various geometrical characteristics of the cells using the potentials introduced in Section \ref{s1}.
For example, in the Voronoi case, the theoretical values (depending on the intensity) for the first two moments of various characteristics can be determined, cf. \cite{Ok}. A simulation study demonstrating that Gibbs-Laguerre tessellations outperform Poisson-Laguerre tessellations in terms 
of the variety of possible cell shapes and characteristics
is presented in the online supplementary material.

\section{Application to polycrystalline microstructures}\label{rd}

This section is devoted to experimental data which are first introduced in Section~\ref{sec:experimentalData}. The main task--how to generate tessellations having similar properties to those of the data--is discussed in Section~\ref{sec:modelSelection}. Here, two approaches are utilized: fitting a parametric model and statistical reconstruction, which are described in the online supplementary material and in Section \ref{ssr}, respectively. Sections \ref{mmt_reco} and \ref{hist_reco} focus on the reconstruction of the data. Two methods of reconstruction are suggested, and their ability to simulate tessellations with prescribed properties is evaluated. 
Section \ref{reco_comp} complements Section \ref{ssr} in the sense of comparison of MHBDM reconstruction and classical method based on greedy algorithm.

\subsection{Experimental data}\label{sec:experimentalData}
\begin{figure}
    \center
    \begin{tikzpicture}
        \node (plots) {\includegraphics[width=0.8\linewidth]{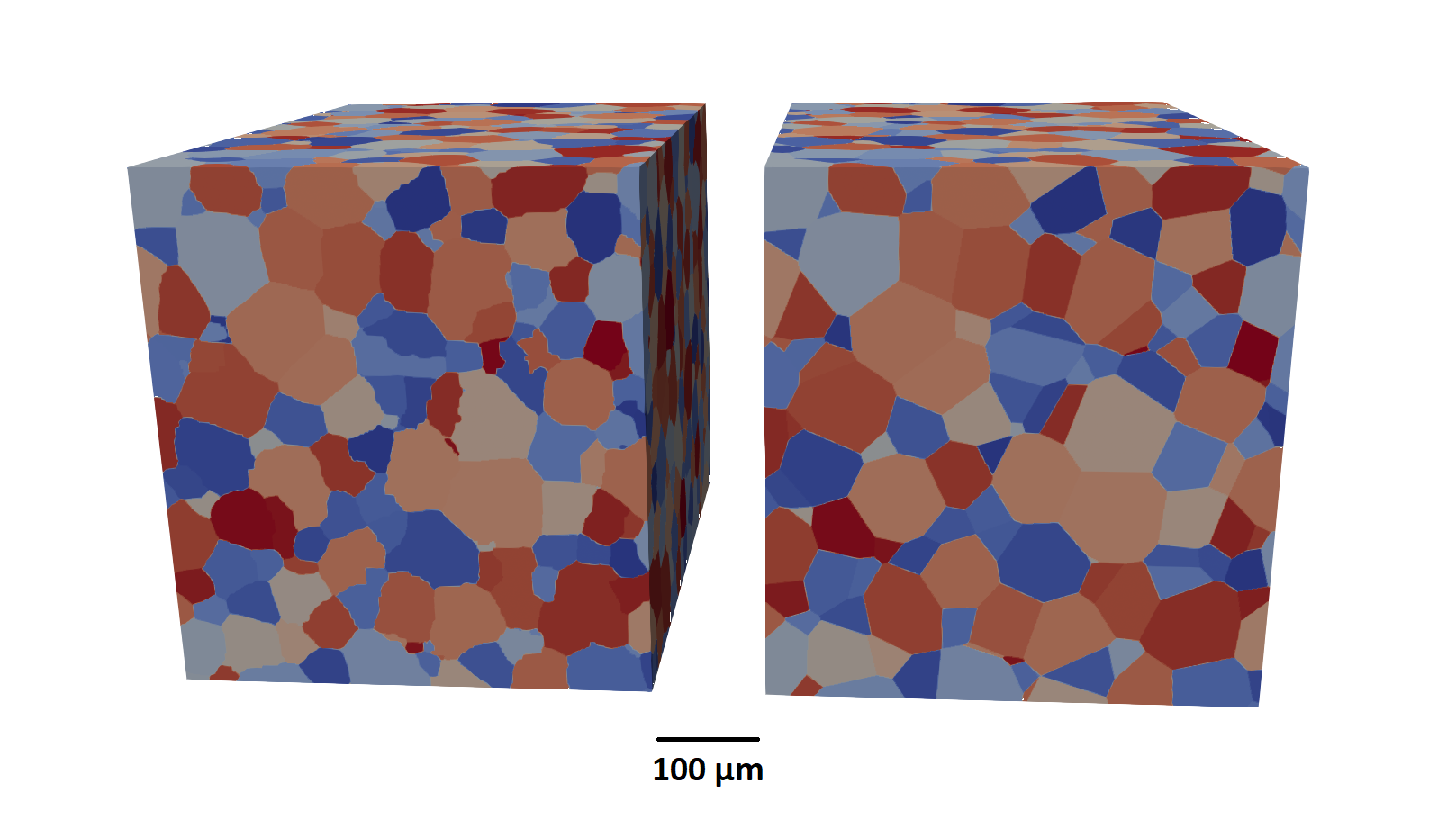}};

        \node[anchor=base, yshift=2ex] at ($(plots.south west)!0.6!(plots.south)$) {a)};
        \node[anchor=base, yshift=2ex] at ($(plots.south)!0.4!(plots.south east)$) {b)};
    \end{tikzpicture}
    \caption{Aluminum alloy specimen measured by synchrotron X-ray tomography (cf. \cite{SP}) --- a cuboid cropped out of the cylindrical domain for the purposes of statistical reconstruction: a) original voxelized image, b) Laguerre approximation serving as experimental dataset}
    \label{expdata}
\end{figure}

The motivation for the simulations described in Section \ref{ss} comes primarily from real experimental data. The image data 
used in this paper are obtained by 
synchrotron X-ray tomographic imaging 
and are a cutout of the polycrystalline microstructure of an Al-5 wt\% Cu sample, which is described in \cite{SP} together with its approximation by a deterministic Laguerre tessellation, see Fig.~\ref{expdata}. 
Later in this section, the Laguerre tessellation from Fig.~\ref{expdata}b is used as our experimental dataset, including its generators. 

In the experimental data there are 
1057 nonempty cells in a cuboidal domain of size $486\times 529\times 685$ $\mu m^3$. The total number of neighboring pairs of cells is 7453. 

\begin{figure}
    \centering
    \includegraphics[width=4cm]{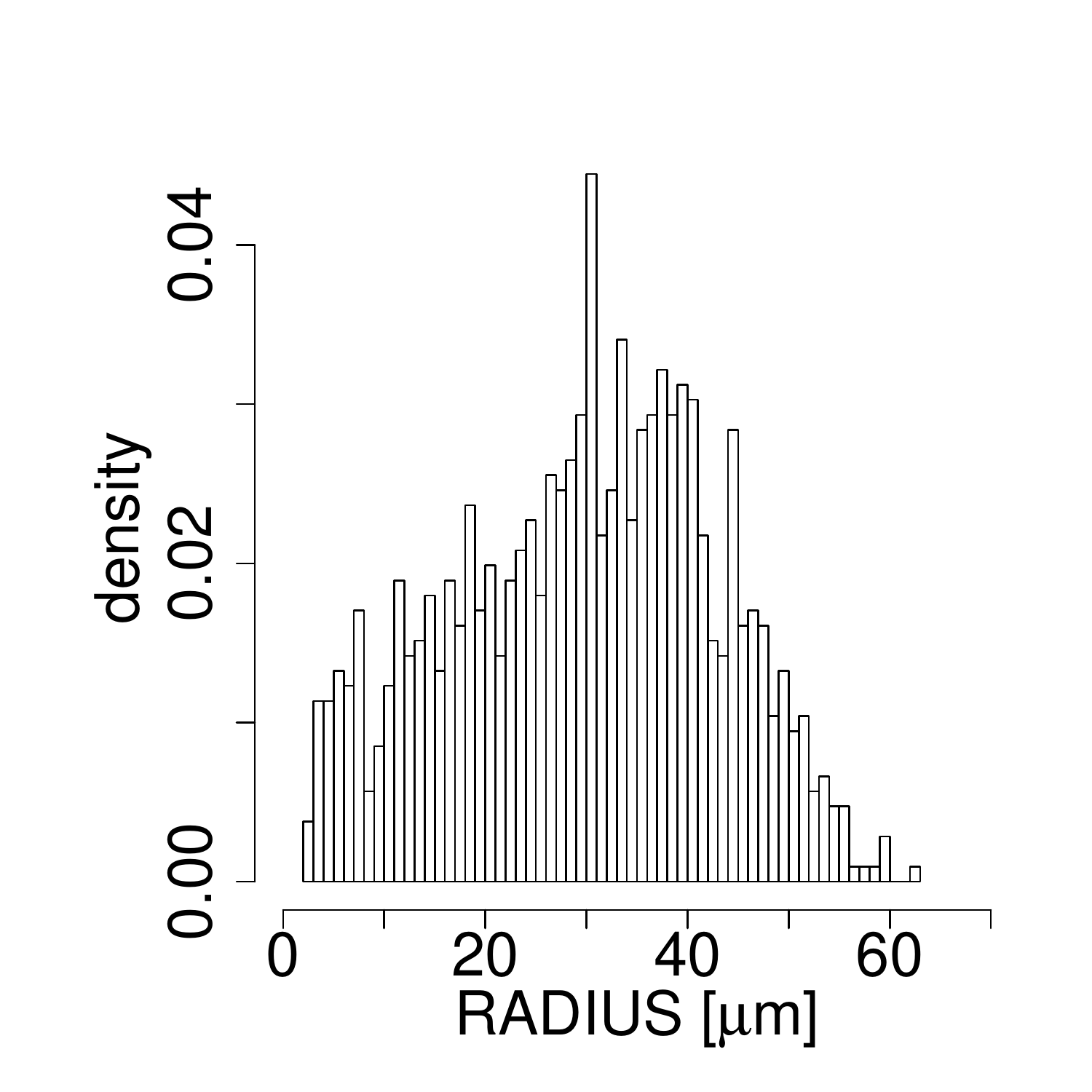}
    \caption{Histogram of relative frequencies of radius marks $[\mu m]$ of the Laguerre tessellation for the experimental data, Fig.~\ref{expdata}b (1034 cells)}
    \label{rads}
\end{figure}

\begin{figure}
    \centering
    \begin{tikzpicture}[node distance=0]
        \node[anchor=north] (plot11) {\includegraphics[width=4cm]{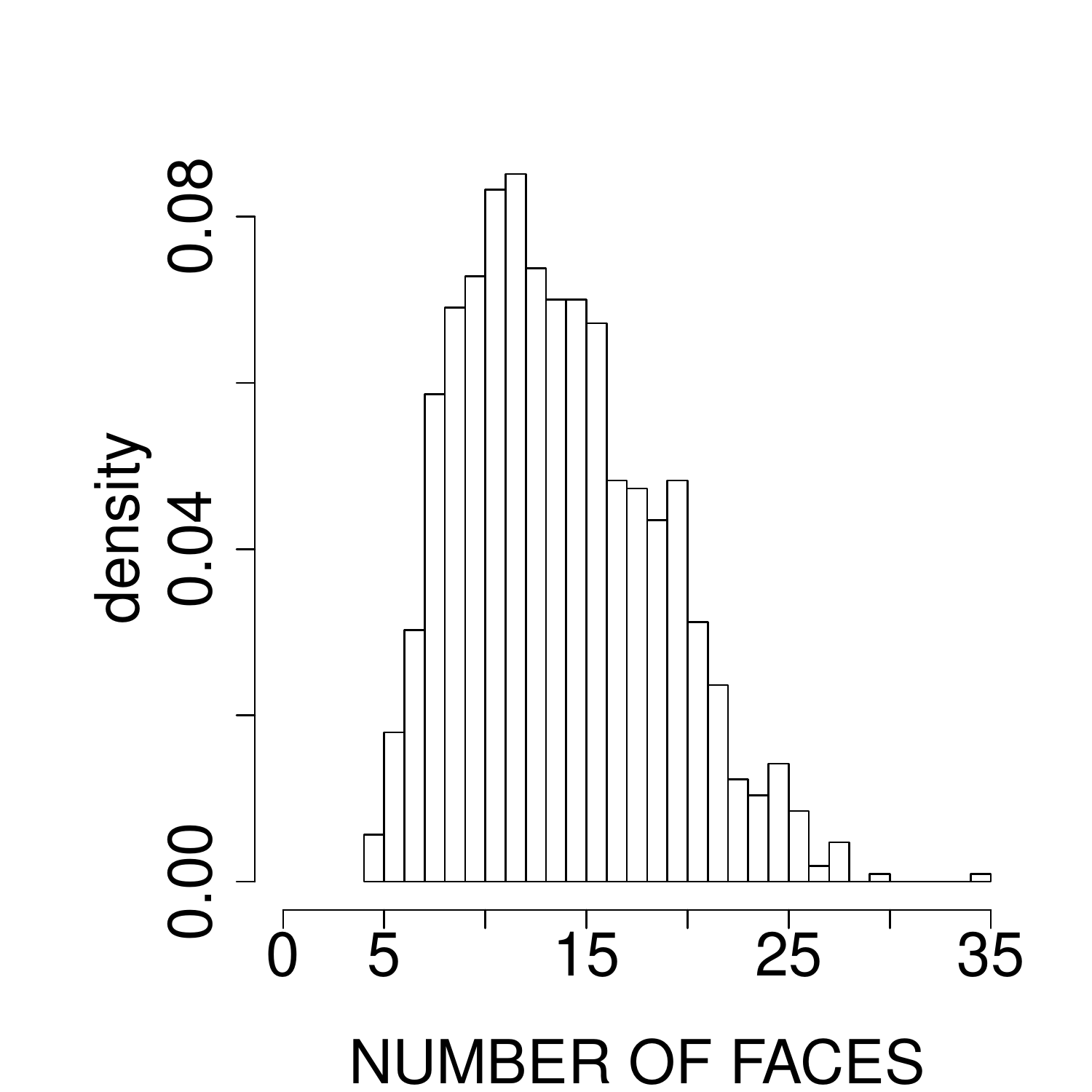}};
        \node[anchor=north, right=of plot11] (plot12) {\includegraphics[width=4cm]{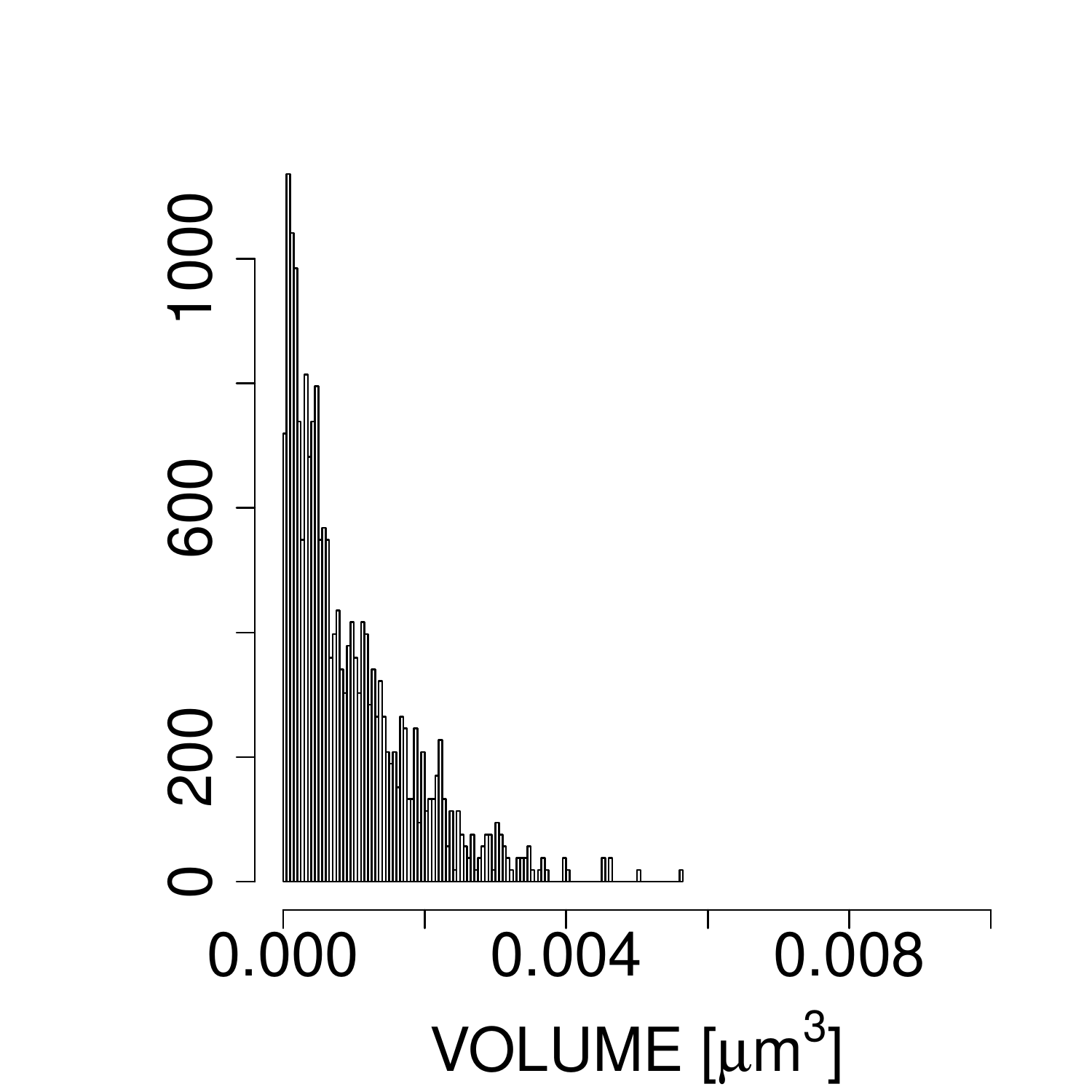}};
        \node[yshift=-2ex] at (plot11.south) {(a)};
        \node[yshift=-2ex] at (plot12.south) {(b)};
        \node[anchor=north, below=of plot11] (plot21) {\includegraphics[width=4cm]{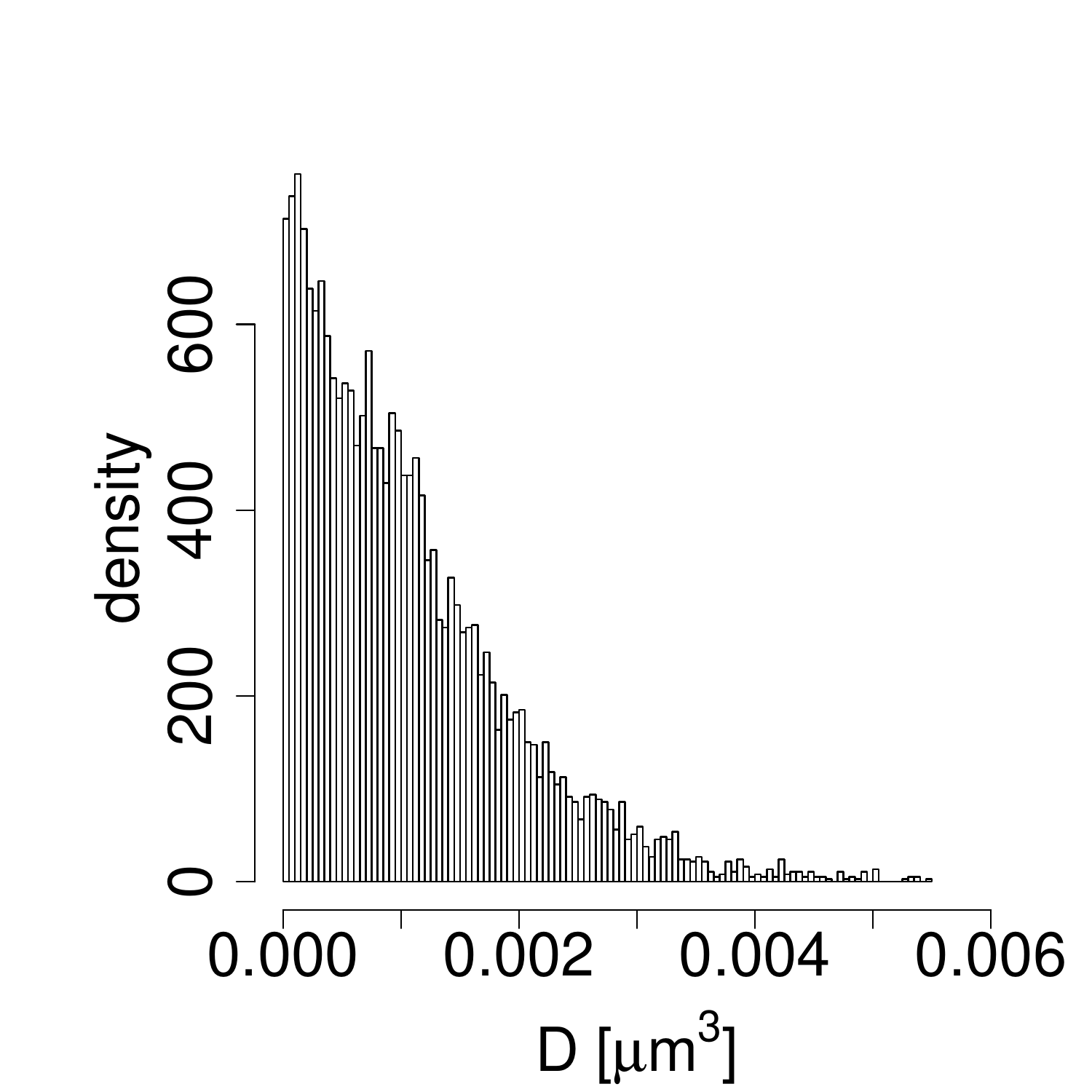}};
        \node[anchor=north, right=of plot21] (plot22) {\includegraphics[width=4cm]{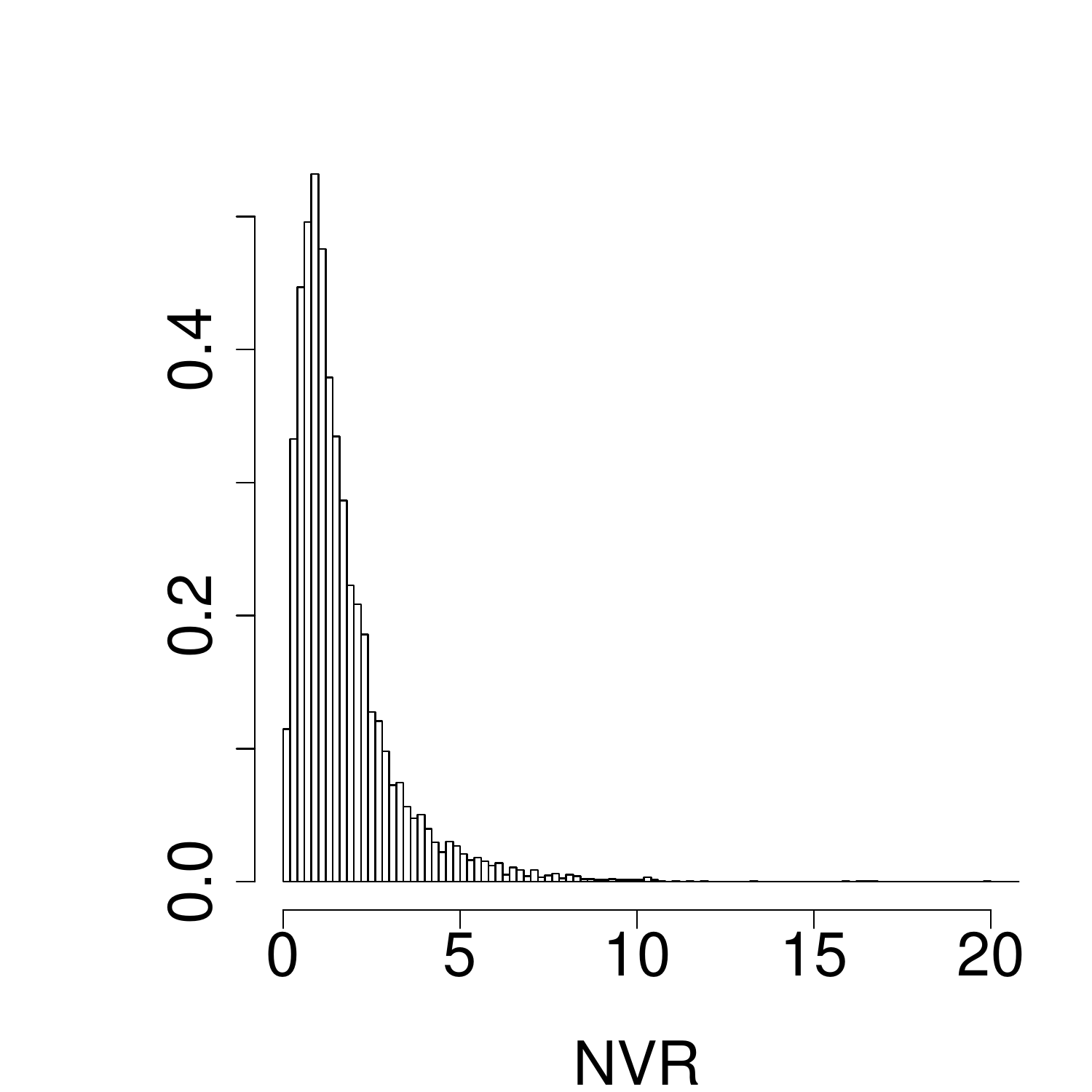}};
        \node[yshift=-2ex] at (plot21.south) {(c)};
        \node[yshift=-2ex] at (plot22.south) {(d)};
    \end{tikzpicture}
    \caption{Normalized geometrical characteristics of the experimental data from Fig.~\ref{expdata}b: histograms of relative frequencies of a number of faces per cell, (a) $hist_{\nof}^{exd}$, the cell volume, (b) $hist_{\vol}^{exd}$, (c) the difference of in cell volumes $[\mu m^3]$, and (d) the neighbor-volume ratio}
    \label{stats}
\end{figure}
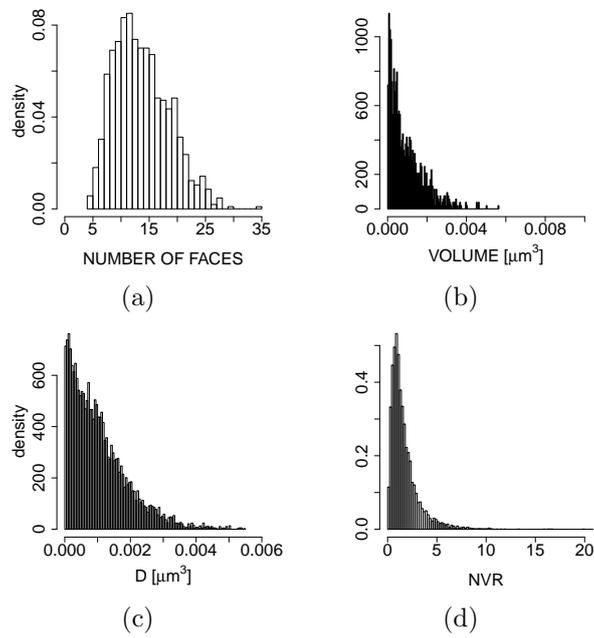

For the purposes of the reconstructions, we normalize to the unit volume. This means that
the volume of each cell is divided by the volume of the cuboidal domain of the experimental data. 
Fig.~\ref{stats} shows histograms of normalized characteristics of the experimental data: namely, the number of faces per cell, the cell volume, the NVR \eqref{pairpo}, and the difference in cell volumes \( D(C_1,C_2) = ||C_1|-|C_2||\) between two neighboring cells $C_1$, $C_2$. Note that comparing Fig.~\ref{stats}d with Fig.~\ref{plot3}, we observe that the NVR of experimental data is closer to an irregular model rather than to a regular one. Table \ref{tabstats} summarizes the moments of the same normalized characteristics.
 Further, the normalization of the radii is a necessary step. As lengths are normalized by the cube root of the volume of the cuboidal domain, the largest radius is below $0.124$ (which corresponds to $70$, cf. Fig.~\ref{rads}, before normalization). Thus, the choice $R_0=0.2$ in Section \ref{sec:BDMA} for the proposal densities in \textit{Algorithm 1} is justified, since 
this value covers all radii of the experimental data.
The activity $z$ is still set to $2000$, because the latter value takes on 
the same order of magnitude as the observed intensity (and the precise value does not have a significant influence on the final intensity, as mentioned above).
The symbols $hist_{\nof}^{exd}$ and $hist_{\vol}^{exd}$ denote the relative histogram of the number of faces in Fig.~\ref{stats}a and the relative histogram of the cell volume in Fig.~\ref{stats}b, respectively. The upper index ``exd" means that the histogram corresponds to experimental data.

\begin{table}
    \centering
    \caption{Statistical description of experimental data: mean and standard deviation of the radius of generators, the number of faces (nof), the volume of cells, the difference of the cell volumes (D), and the neighbor-volume ratio}
    \label{tabstats}
    \begin{tabular}{llllll}
        \hline\noalign{\smallskip}
                                                       & Radius $[\mu m]$  & nof     & Volume $[\mu m^3]$   & D $[\mu m^3]$        & NVR \\
        \noalign{\smallskip}\hline\noalign{\smallskip}
        Mean                                          & 29.7693 & 14.1608 & $9.6712\cdot10^{-4}$ & $1.0703\cdot10^{-3}$ & 1.6995 \\
        SD                                             & 18.9564 & 4.8558  & $1.0782\cdot10^{-4}$ & $8.9165\cdot10^{-4}$ & 1.8754 \\
        \noalign{\smallskip}\hline
    \end{tabular}
\end{table}

\subsection{Modeling approach}\label{sec:modelSelection}
The aim is to create models of random tessellations whose realizations are similar to the experimental data. The first approach to doing so it to estimate the parameter values of various Gibbs point processes chosen a priori, using standard techniques like the pseudolikelihood method, which is described in the online supplementary material. 
Some practical aspects regarding the problem of estimation by the maximum pseudolikelihood method are mentioned in \cite{DL}, Section 4 (estimation of the parameters $\theta$ and $z$ of the Gibbs-Voronoi model is commented on in Section A.3). 
An alternative approach preferred here is the statistical reconstruction method described in Section~\ref{ssr}. The rest of this section applies the latter method to two examples.

\subsection{Moment reconstruction}\label{mmt_reco}
First, we aim to reconstruct the experimental data using moments. In the online supplementary material we present simulations that force the realizations to have a prescribed average number of faces per cell (class of random tessellations $RT3$). These simulations are successful in the sense that they match the prescribed value. On the other hand, the variance and overall shape of the distribution can be entirely different even within a single specification. 
Therefore, we investigate the first and second moments together. This is carried out using the random tessellation class $RT4$, considering either solely the number of faces per cell or solely the cell volume. Moreover, both of these geometrical characteristics can be considered together. The energy function of the Gibbs-Laguerre model consists of four potentials in random tessellation $RT5$. With an increasing number of potentials combined in the energy function, it becomes more and more difficult to match the prescribed values, but even for the four potentials of the random tessellation class $RT5$ the results are satisfactory.
Numerical results for the random tessellations $RT4$ and $RT5$ can be found in the online supplementary material.

\subsection{Histogram reconstruction}\label{hist_reco}
A more sophisticated approach to the statistical reconstruction of tessellations is to control not only a few moments but the entire distribution of a geometrical characteristic. The easiest way to accomplish this is to measure the discrepancy between histograms, see Fig.~\ref{fig:discrepancy}. We will consider two different setups. The first one controls the distribution of the number of faces per cell, and the second one adds the distribution of the cell volumes. In addition to providing the results, we examine the choice of parameters in detail. The following parameter specifications are used:
\begin{equation}   \label{model8}
s=\nof, \quad T(s)=\dsc(H_s,H_s^{\prime}),\quad  H_{\nof}^{\prime}=hist_{\nof}^{exd}, \quad \theta_n, \\
\end{equation}
(denoted in Table \ref{Tno} as $RT6$)
and
\begin{equation}   \label{model10}
\begin{array}{l}
s_1=\nof, \quad T_1(s)=\dsc(H_s,H_s^{\prime}),\quad  H_{\nof}^{\prime}=hist_{\nof}^{exd}, \quad \theta^1_n, \\
s_2=\vol, \quad T_2(s)=\dsc(H_s,H_s^{\prime}),\quad  H_{\vol}^{\prime}=hist_{\vol}^{exd}, \quad \theta^2_n, \\
\end{array}
\end{equation} (denoted in Table \ref{Tno} as $RT7$).
Since the parameters $\theta_n, \theta_n^1, \theta_n^2$ are unspecified, $RT6$ and $RT7$ form classes of tessellations. Once again, the stopping criterion employed is $(\delta, t) = (0.002, 500\,000)$. 

\begin{table}
    \centering
    \caption{Dependence on $\theta_n$ of the discrepancy between histograms for the number of faces per cell and for the cell volumes for tessellations of the class $RT6$}
    \label{tab5}
    \begin{tabular}{ccc}
        \hline\noalign{\smallskip}
        $\theta_n$ &  \multicolumn{2}{c}{discrepancy} \\
        &  nof & volume \\
        \noalign{\smallskip}\hline\noalign{\smallskip}
        10          & 0.46822 & 0.70841 \\
        100         & 0.41481 & 0.68919 \\
        1\,000      & 0.02964 & 0.45739 \\
        10\,000     & 0.02529 & 1.32095 \\
        100\,000    & 0.02356 & 1.28102 \\
        1\,000\,000 & 0.02102 & 1.32543 \\
        \noalign{\smallskip}\hline
    \end{tabular}
\end{table}

\begin{figure}
\centering
    \begin{tikzpicture}[node distance=-1cm]
        \node (plot1) {\includegraphics[width=0.4\linewidth]{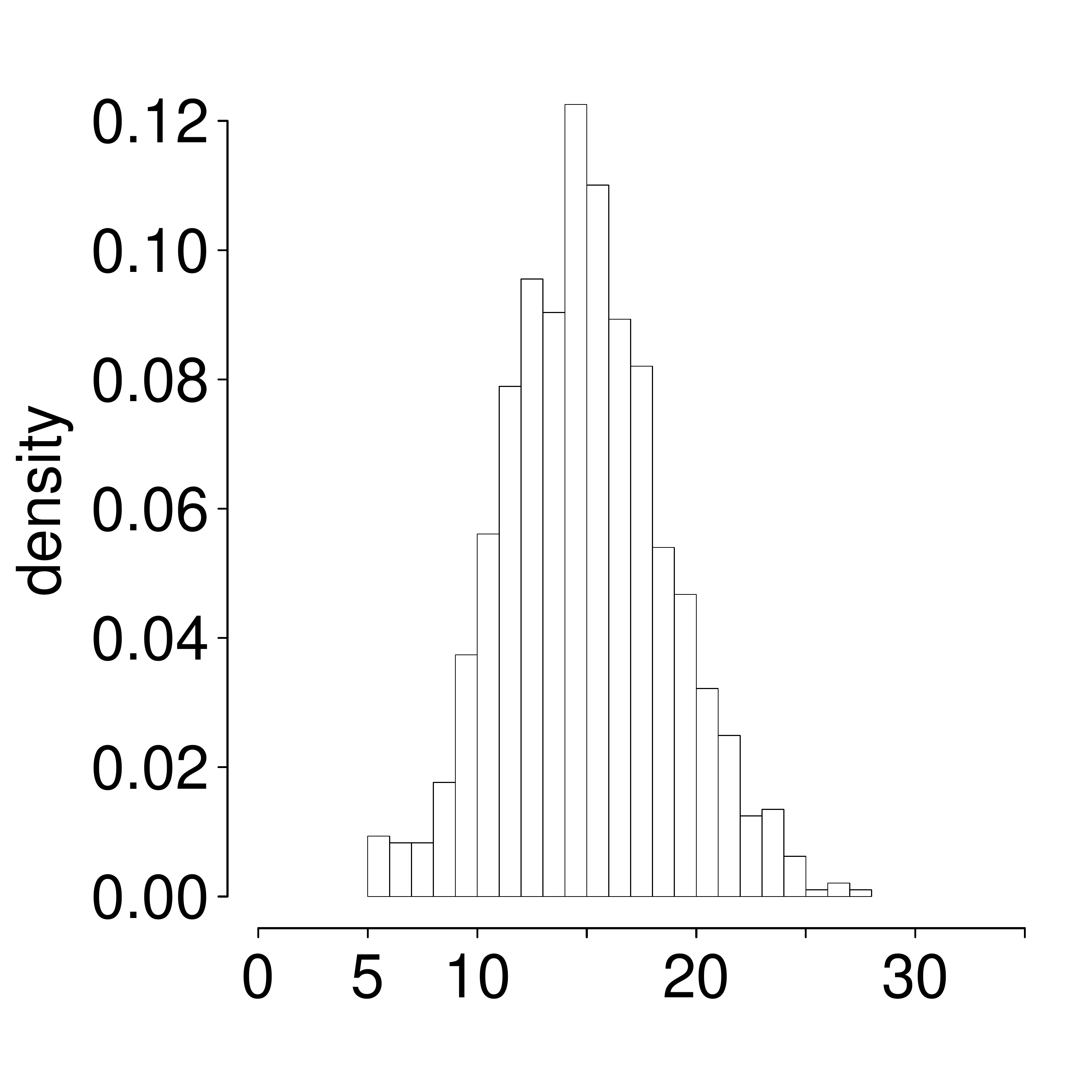}
        \includegraphics[width=0.4\linewidth]{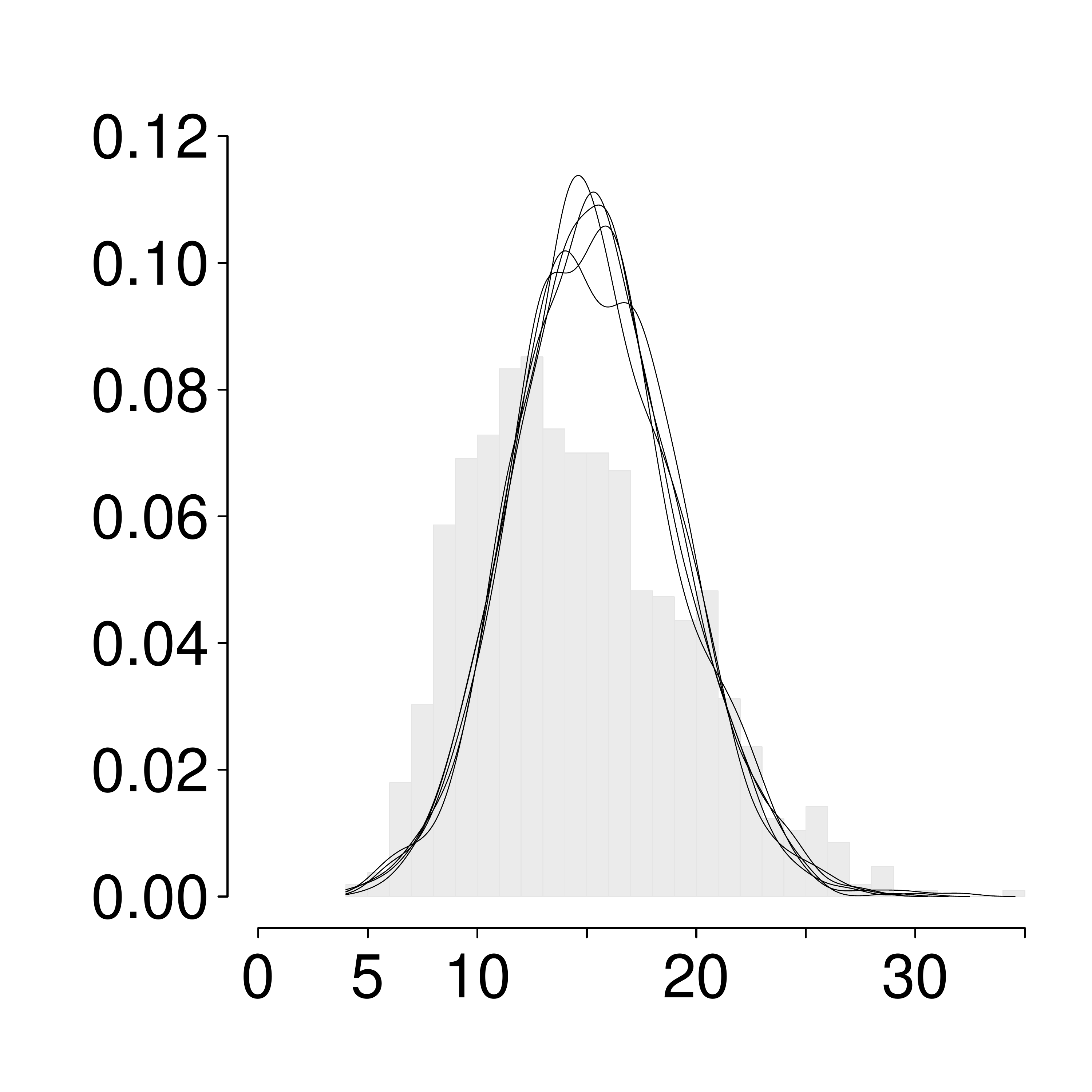}};
        \node[below=of plot1] (plot2) {\includegraphics[width=0.4\linewidth]{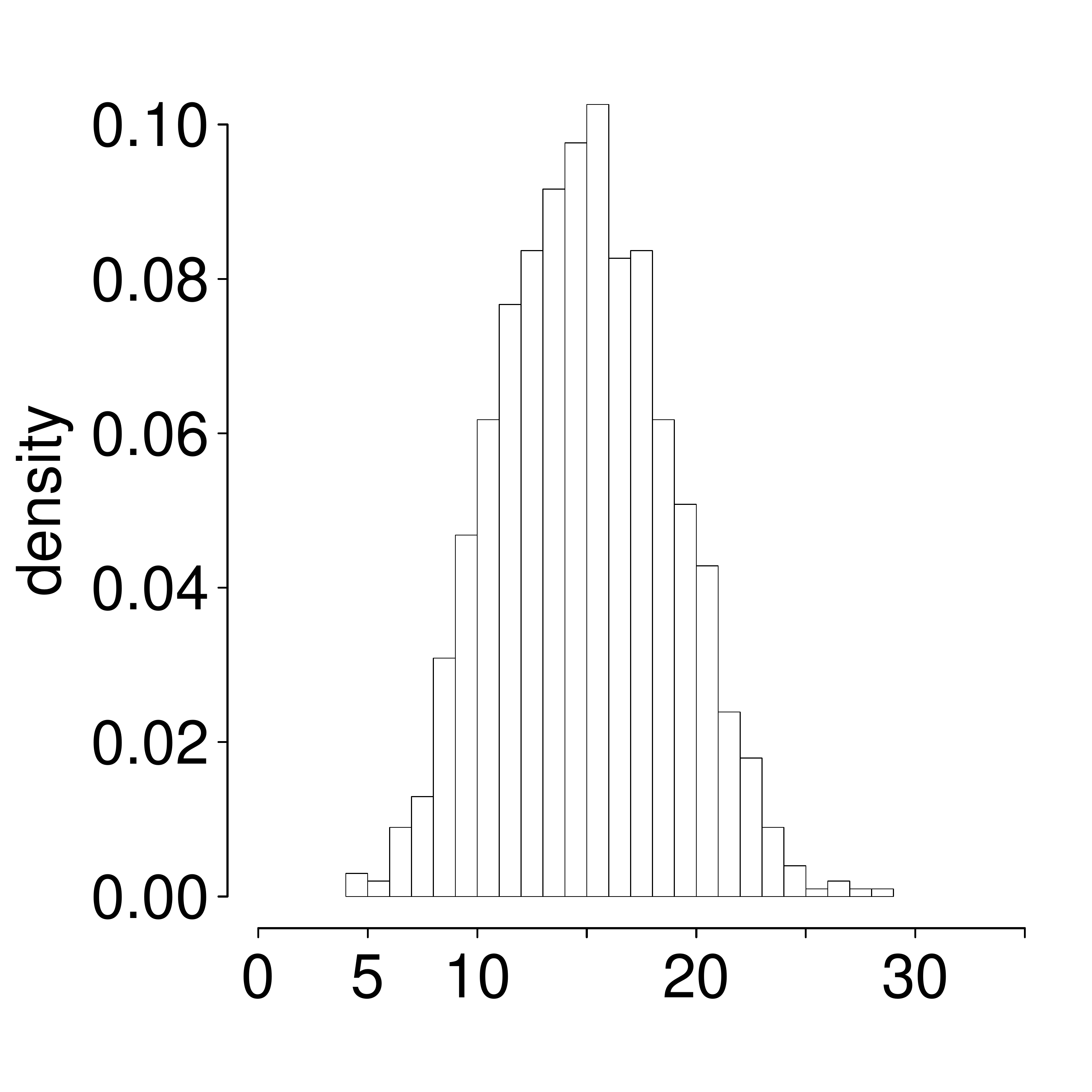}
        \includegraphics[width=0.4\linewidth]{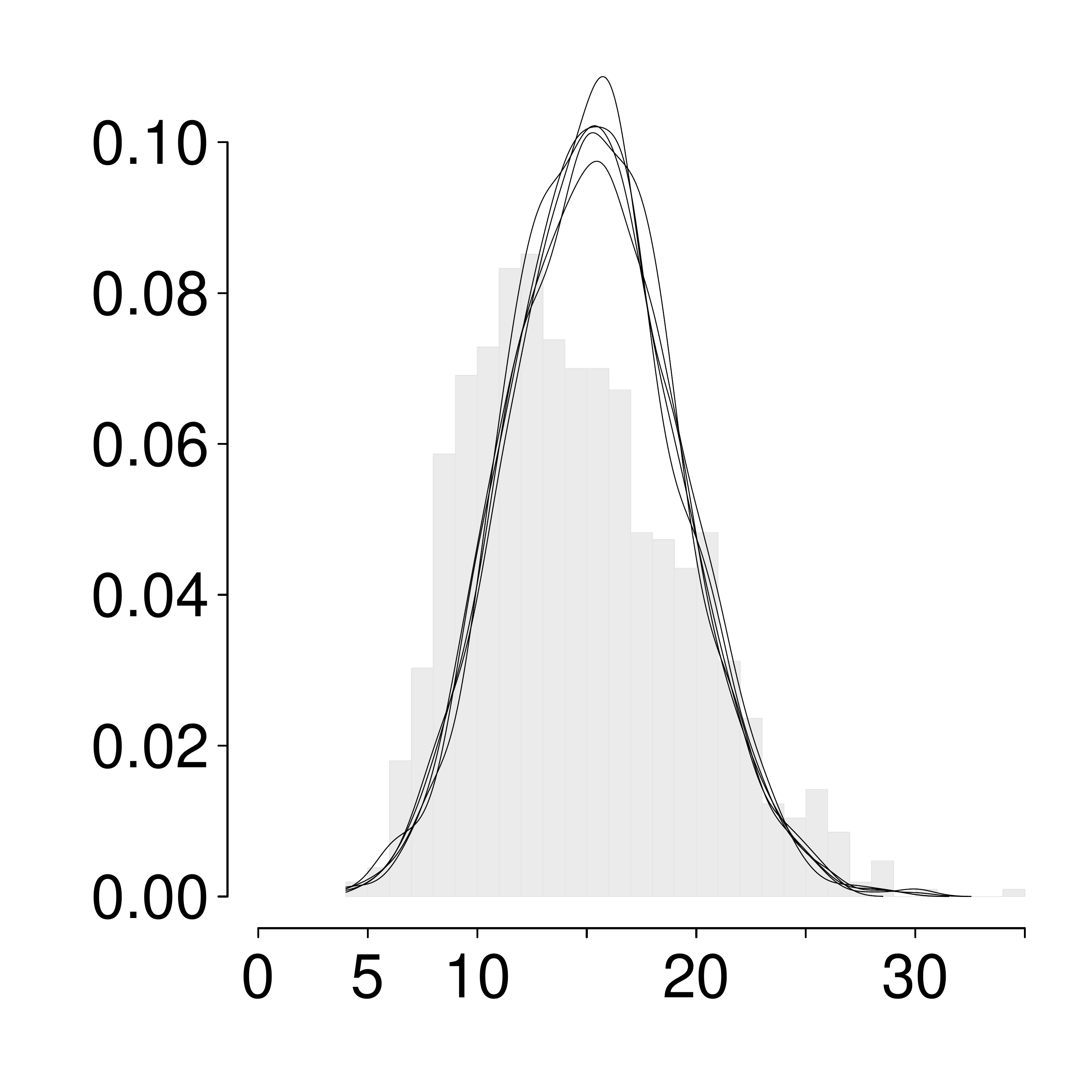}};
        \node[below=of plot2] (plot3) {\includegraphics[width=0.4\linewidth]{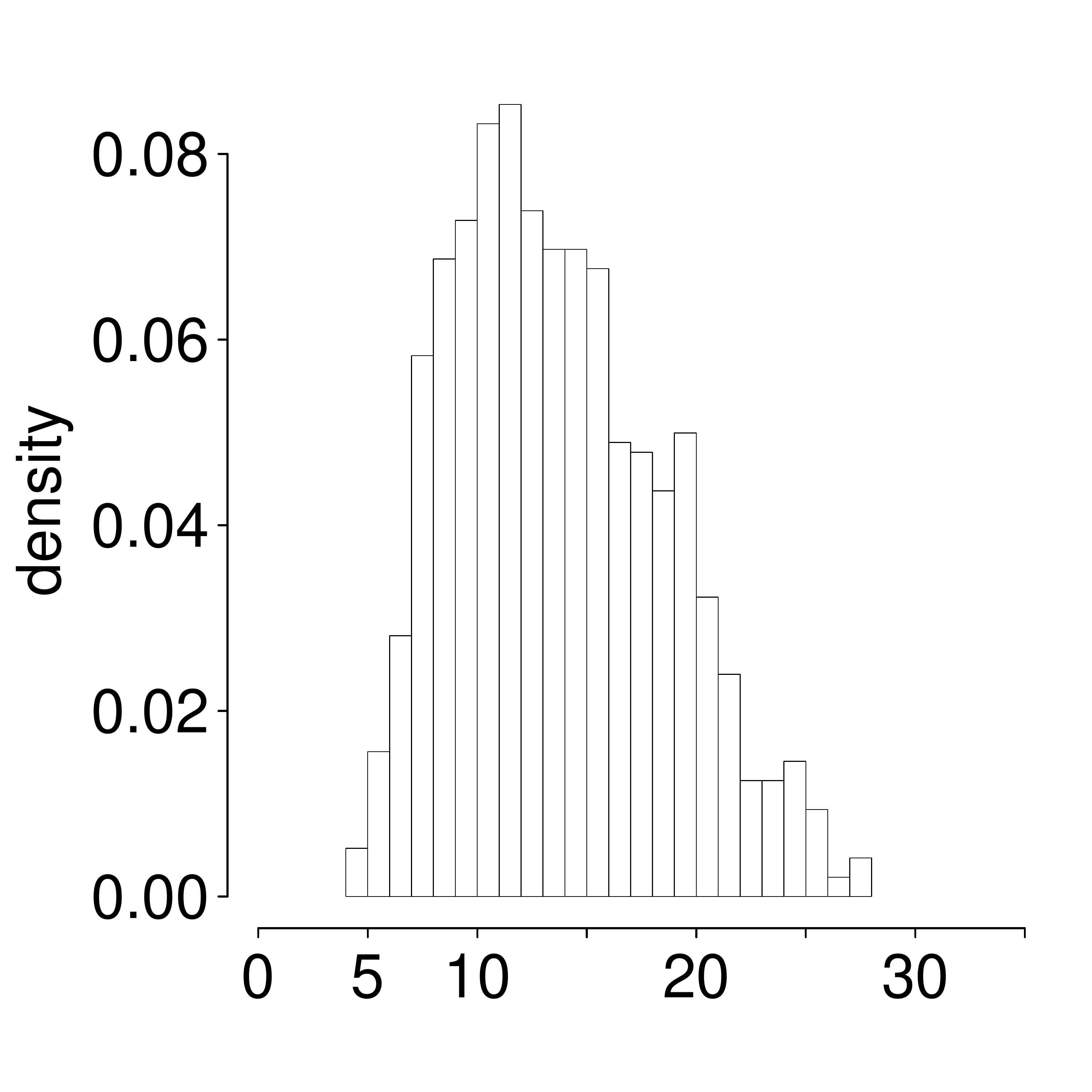}
        \includegraphics[width=0.4\linewidth]{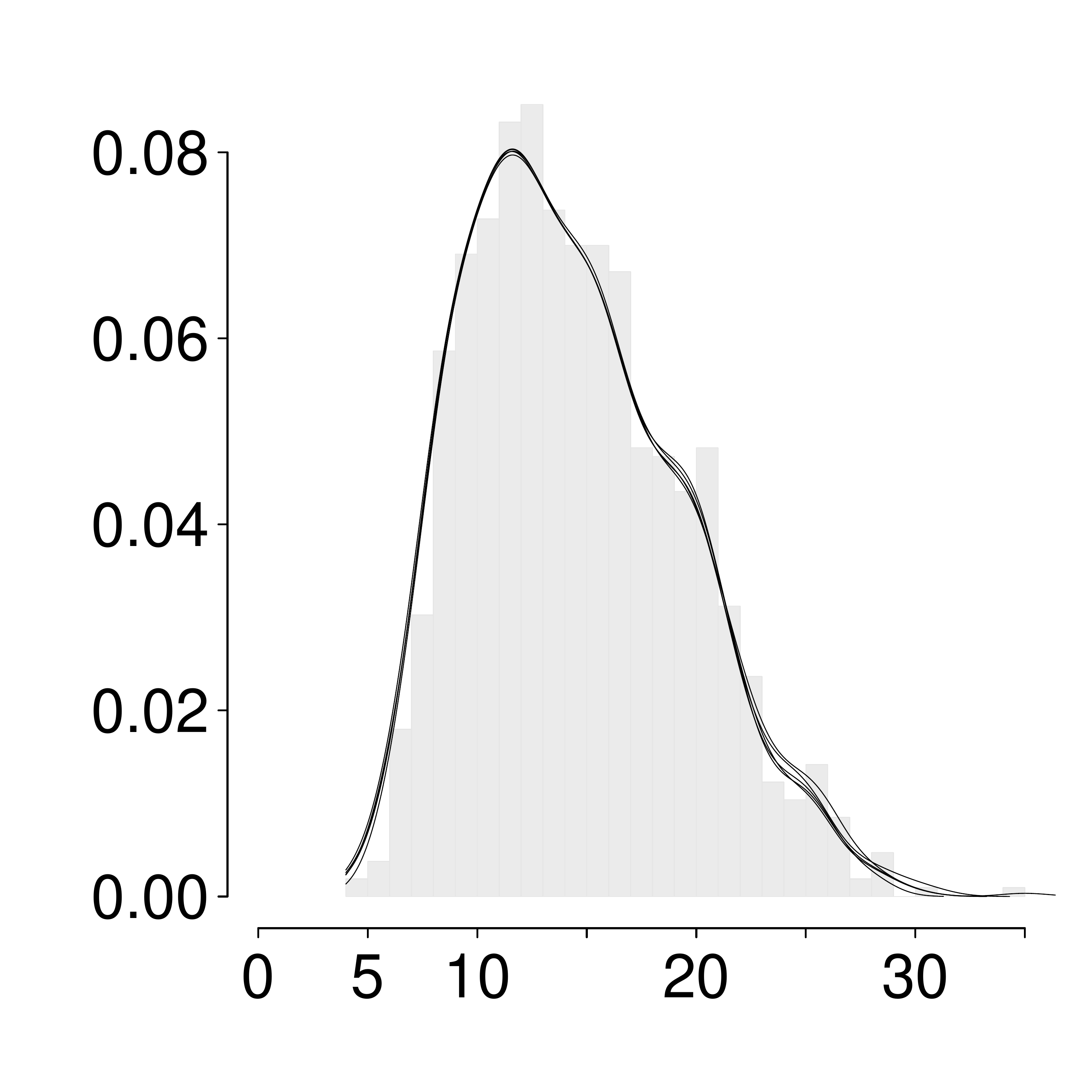}};
        \node[below=of plot3] (plot4) {\includegraphics[width=0.4\linewidth]{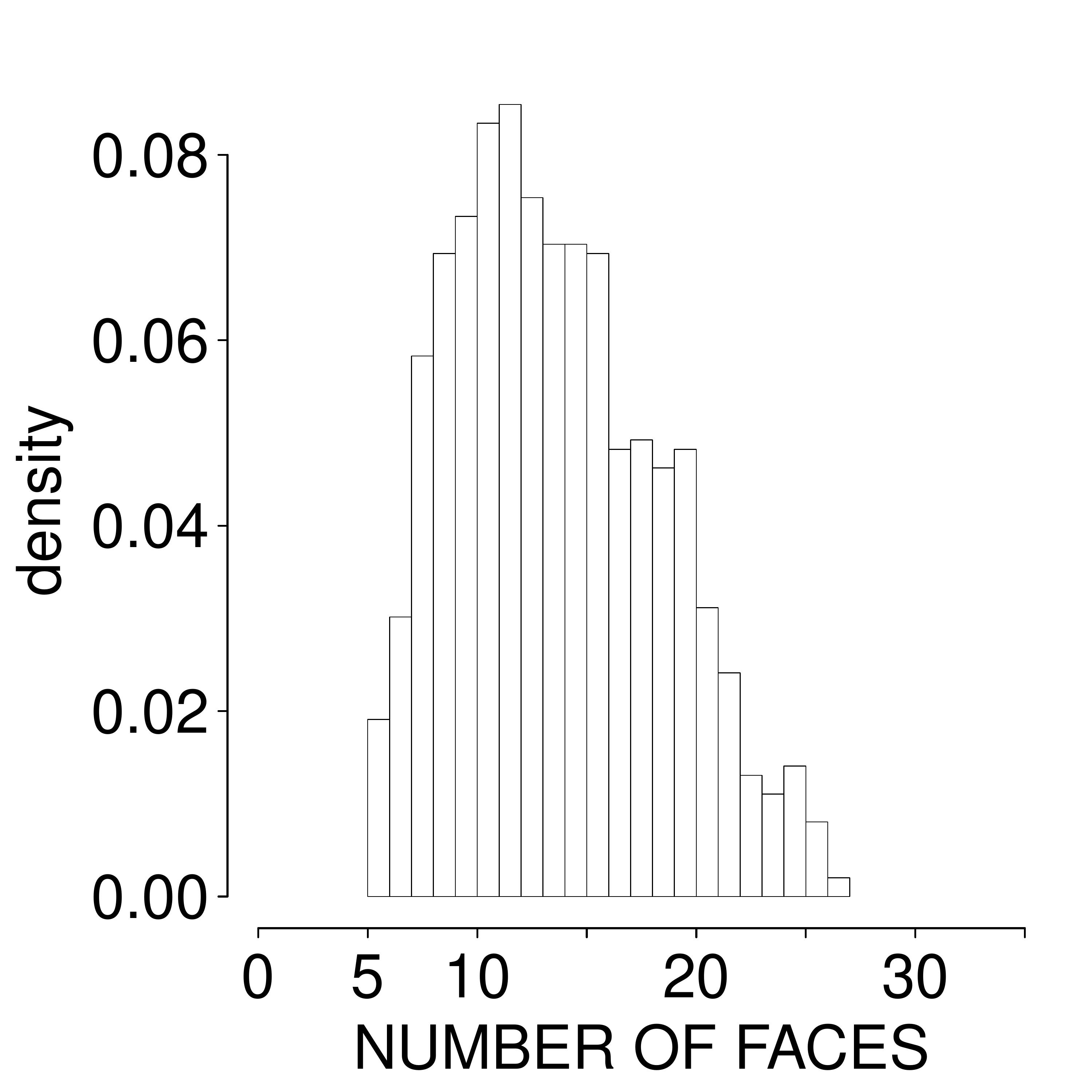}
        \includegraphics[width=0.4\linewidth]{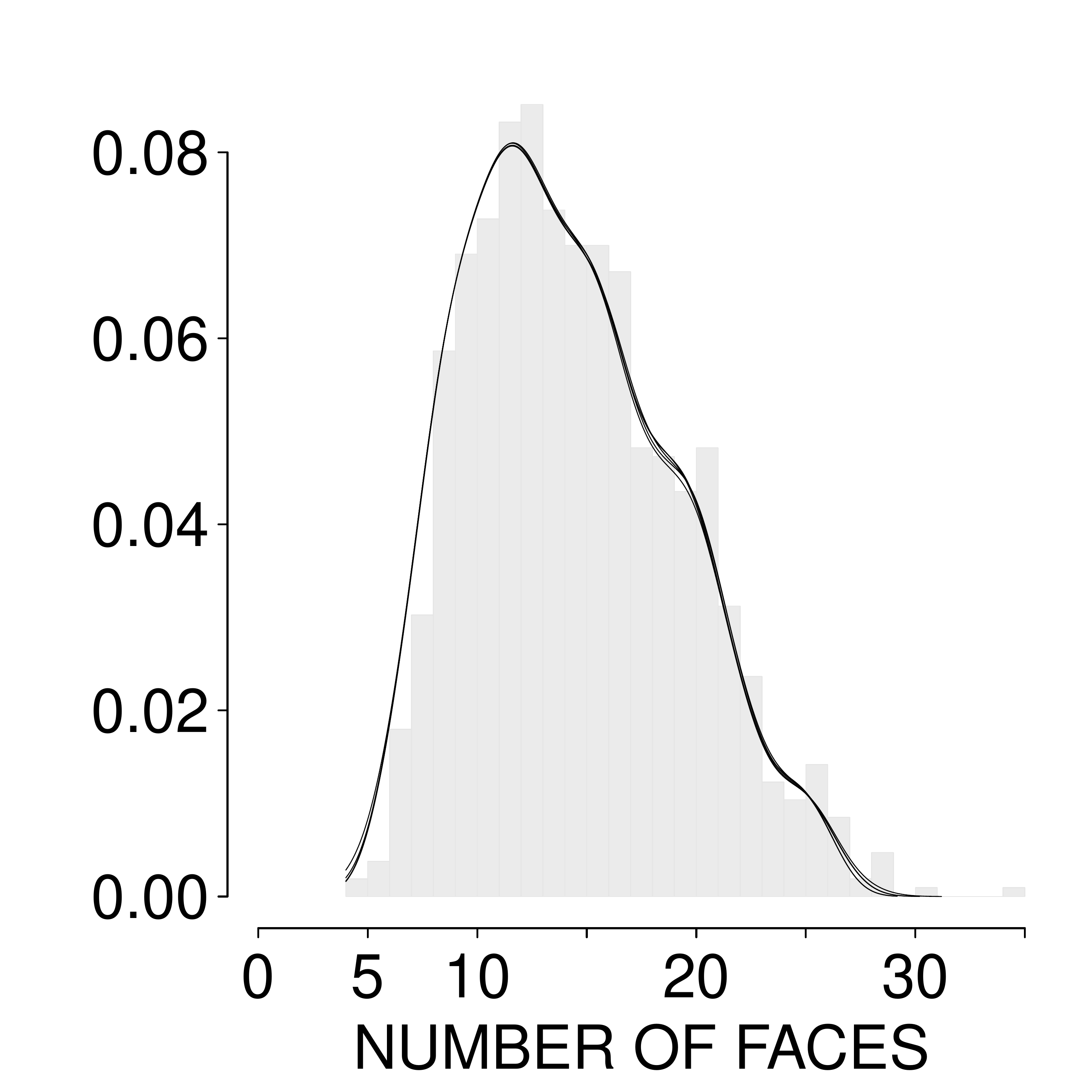}};

        \node at ($(plot1.north west)!0.5!(plot1.north)$) {A};
        \node at ($(plot1.north)!0.5!(plot1.north east)$) {B};
    \end{tikzpicture}
    \caption{
Reconstruction of experimental data, controlling the distribution of the numbers of faces per cell (class $RT6$, \eqref{model8}): from top to bottom the parameter $\theta_n$ takes on the values $10$, $100$, $1\,000$ and $10\,000$; column A shows the histograms of relative frequency computed from a single realization, and column B shows kernel density estimates based on ten realizations together with the histogram coming from the experimental data (cf. Fig.~\ref{stats}a)---in gray
}
    \label{plot8}
\end{figure}

Fig.~\ref{plot8} shows how the value of $\theta_n$ influences the variability of the simulated realizations. The results should be compared to the experimental data, see Fig.~\ref{stats}a. Table \ref{tab5} demonstrates what happens with the discrepancy when $\theta_n$ increases. The reconstructing potential considers only the number of faces per cell ($\nof$); therefore, the discrepancy of the histograms of cell volume is not controlled. In summary, a small value of $\theta_n$ results in a large discrepancy for histograms of the number of faces per cell. On the other hand, increasing $\theta_n$ beyond a certain level leads to no further improvement, because the acceptance ratios tend to zero
(see Section \ref{control}). The variability of the realizations is higher for small values of $\theta_n$ and it decreases when $\theta_n$ grows.  In the online supplementary material we present another class of random tessellations based on a single potential concerning the histogram of cell volumes. It is an analogy of the class $RT6$, and similar behavior can be observed when changing the value of parameter $\theta_n$. 

\begin{figure}
    \begin{tikzpicture}
        \node (plot) {\includegraphics[width=5cm]{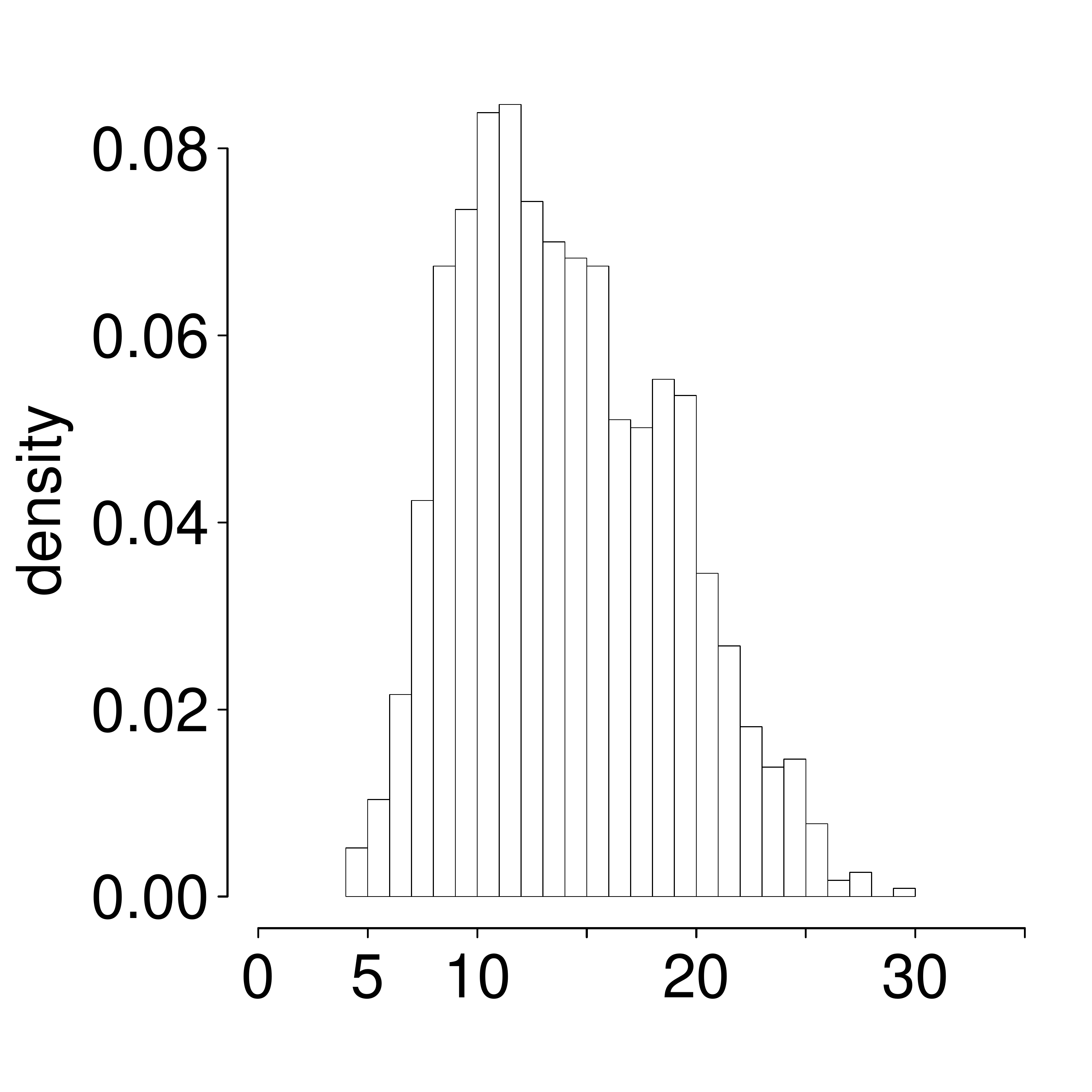}
        \includegraphics[width=5cm]{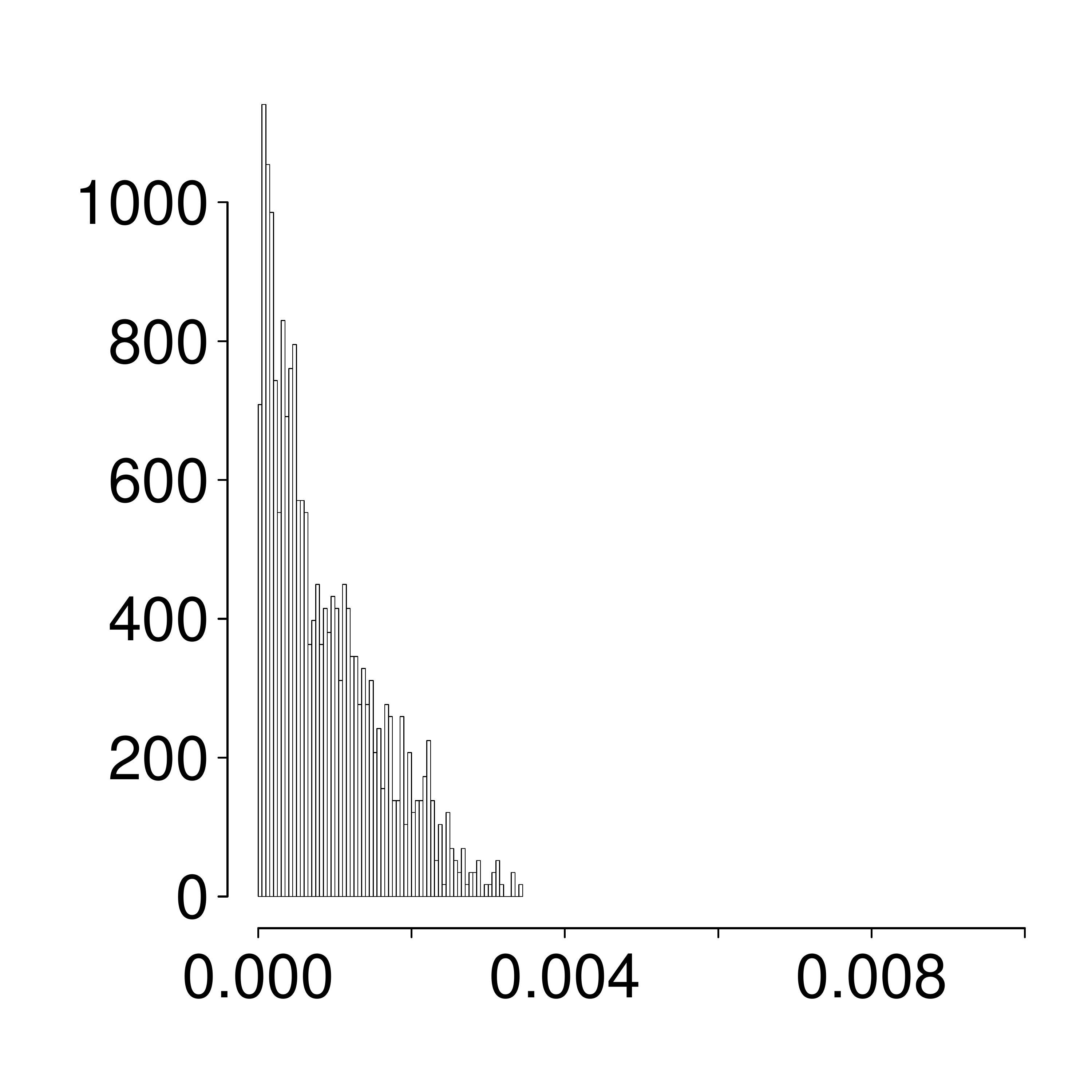}};
        \node at ($(plot.north west)!0.5!(plot.north)$) {A};
        \node at ($(plot.north)!0.5!(plot.north east)$) {B};
        
    \end{tikzpicture}
    \begin{tikzpicture}
        \node (plot) {\includegraphics[width=5cm]{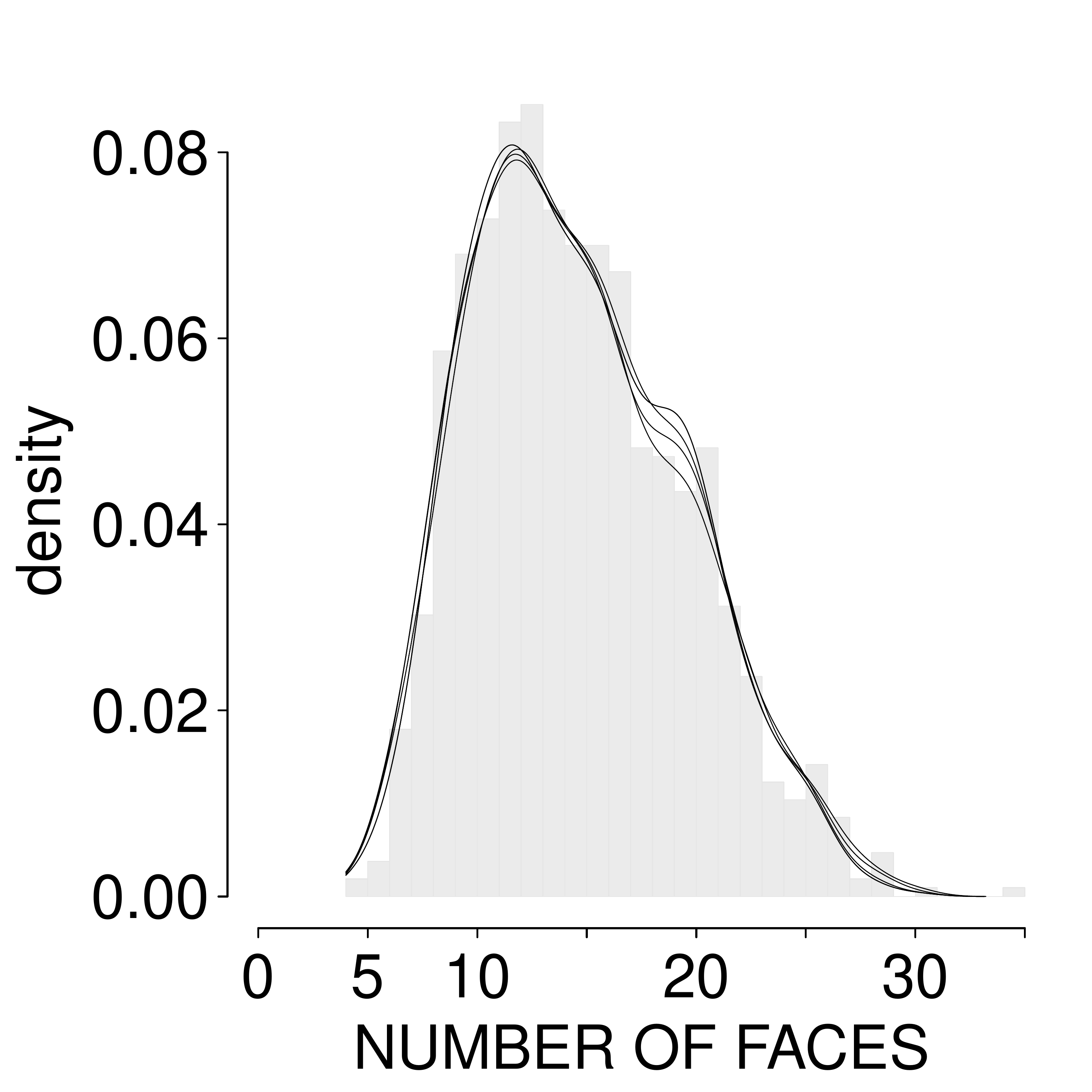}
        \includegraphics[width=5cm]{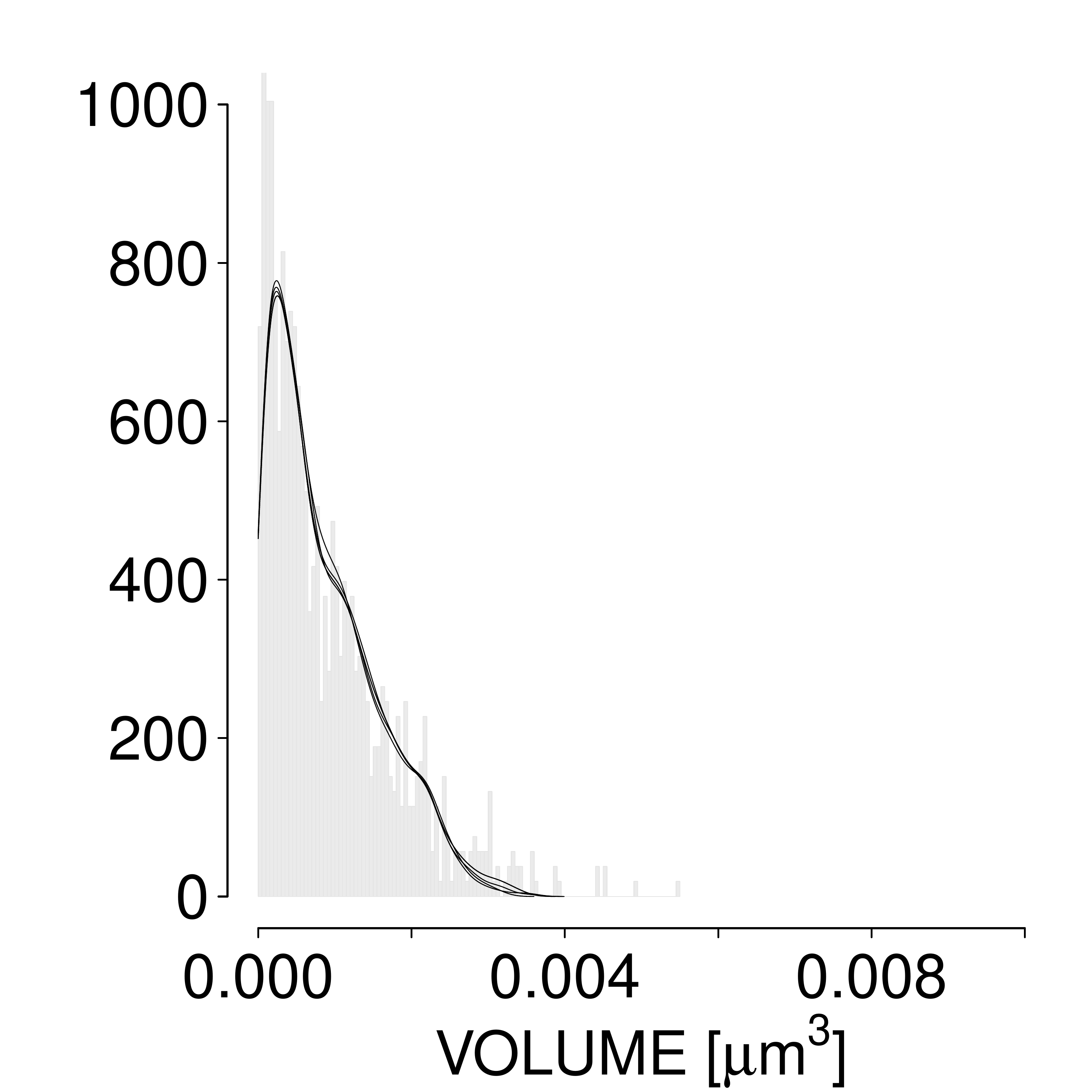}};
        
    \end{tikzpicture}
    \caption{Reconstruction of experimental data, controlling distributions of the number of faces per cell and of cell volume,  class $RT7$, \eqref{model10}) with $\theta_n^1 = 1\,000$ and $\theta_n^2 = 10\,000$: column A shows the number of faces per cell, the histogram of relative frequencies computed from one realization and kernel density estimates based on ten realizations; column B shows the same plots for the cell volume. The gray histograms are those of the experimental data, cf. Fig.~\ref{stats}a,b
}
    \label{plot10}
\end{figure}

The conclusions from the last paragraph remain valid even for random tessellations  from class $RT7$, \eqref{model10}, which combine two potentials based on the histogram discrepancy. Combining more than one potential introduces some difficulties. The values of both parameters have to be in a reasonable proportion as described in Table \ref{tab6}. 
Moreover, it is easy to see that the value of the parameter corresponding to the histogram of cell volumes must be the larger of the two. Fig.~\ref{plot10} shows the reconstruction results for the tessellations from the class $RT7$ in the case of $\theta_n^1 = 1\,000$ and $\theta_n^2 = 10\,000$. The results should be compared to the experimental data, see Fig.~\ref{stats}, in order to verify the success of the reconstruction visually.

\begin{table}
\centering
\caption{Dependence on the parameters $theta_n^1$ and $theta_n^2$ of the discrepancy of histograms for the number of faces per cell and for the cell volume for tessellations of the class $RT7$, \eqref{model10}}
\label{tab6}
\begin{tabular}{cc|cccc}
\hline\noalign{\smallskip}
$\theta_n^1 \backslash \theta_n^2$ & & 1\,000 & 10\,000 & 100\,000 & 1mil \\
\noalign{\smallskip}\hline\noalign{\smallskip}
 \multirow{2}{*}{100}      & nof    & 0.28966            & 0.21418            & 0.20543 & \multirow{2}{*}{-} \\
				           & volume & 0.36386            & 0.06484            & 0.05571 & \\
 \multirow{2}{*}{1\,000}   & nof    & 0.05294            & 0.07903            & 0.14432 & 0.13548 \\
				           & volume & 1.21265            & 0.08136            & 0.06971 & 0.06634 \\
 \multirow{2}{*}{10\,000}  & nof    & \multirow{2}{*}{-} & 0.01671            & 0.06802 & 0.09268 \\
				           & volume &                    & 1.14779            & 0.09701 & 0.06514 \\
 \multirow{2}{*}{100\,000} & nof    & \multirow{2}{*}{-} & \multirow{2}{*}{-} & 0.01327 & 0.05756 \\
				           & volume &                    &                    & 1.02774 & 0.09387 \\
\noalign{\smallskip}\hline
\end{tabular}
\end{table}

\medskip
\medskip
\medskip

\subsection{Comparison of reconstruction approaches}\label{reco_comp}

In this section, a short comparative study is presented of reconstruction via MHBDM (introduced in Section 3.3) and the classical approach using the greedy algorithm described in [5], Chapter 16.
The following pseudocode briefly describes the greedy approach to statistical reconstruction. 


\paragraph{Algorithm 4} (Reconstruction via greedy algorithm). 
\begin{enumerate}
    \item construct an admissible marked point configuration $\x_0$ of $M$ marked points that generate only nonempty Laguerre cells, 
    \item choose a point $(y,r)$ from $\x_0$ at random, generate $(x,s) \sim U_{\x_0 \setminus \{(y,r)\}}(\Lambda \times I)$ and set
                \[ \x_1= (\x_0\setminus\{(y,r)\})\cup \{(x,s)\}, \]
    \item $\x_0 \leftarrow \x_1$ if $\tilde{E}(\x_1) < \tilde{E}(\x_0)$,
    \item if the point configuration $\x_0$ has not changed over the last $L$ iterations, then return $\x_0$, else goto 2. 
\end{enumerate}

The periodic energy, cf. (\ref{parametric}), in step 3 is of the form \eqref{energy_r} and will be specified later. The reconstruction starts by fixing the total number $M$ of nonempty cells in the sampling window $\Lambda \times I$. Then an admissible marked point pattern $\x_0$ of $M$ generators is sampled uniformly in $\Lambda \times I$ such that it generates only nonempty Laguerre cells. In each iteration of \textit{Algorithm 4} a random marked point $(y,r) \in \x_0$ is chosen and proposed to be replaced by a new marked point $(x,s)$. Here, the marked point $(x,s)$ is generated uniformly on the subset of $\Lambda \times I$ that ensures that the corresponding cell is nonempty, i.e., on
$\{(z,u) \in \Lambda\times I: \text{ the cell generated by }(z,u)\text{ is nonempty}\}$. This set depends on the marked point configuration $\x_0 \setminus \{(y,r)\}$ and the appropriate conditional uniform distribution, which we denote by $U_{\x_0 \setminus \{(y,r)\}}(\Lambda \times I)$. The replacement is carried out if the periodic energy of the proposal $\x_1$ is smaller than the periodic energy of $\x_0$. The reconstruction ends if there is no replacement in $L\in\N$ consecutive iterations.

\begin{figure}
\centering
    \begin{tikzpicture}[node distance=0.5cm]
        \node (plot1) {\includegraphics[width=0.4\linewidth]{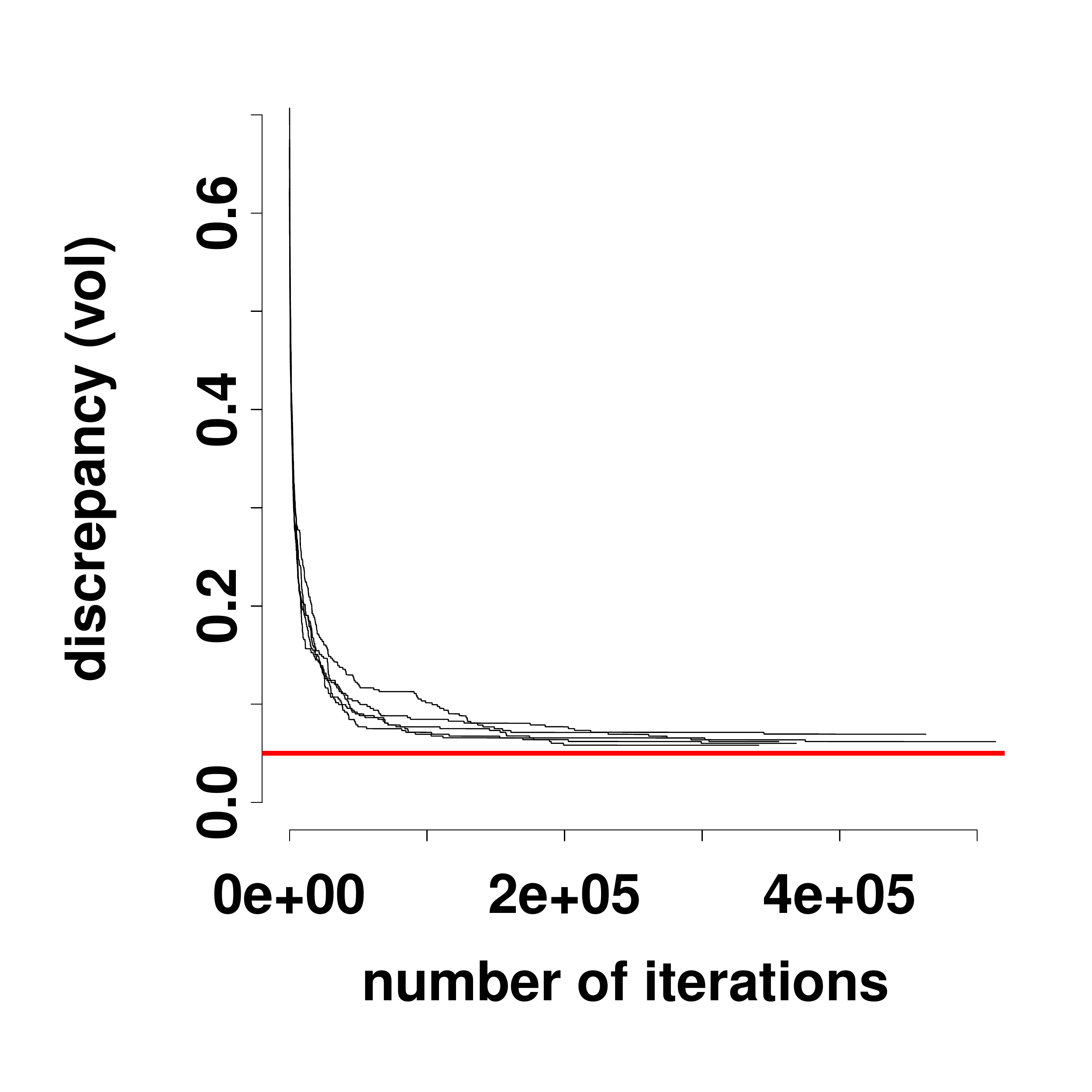}};
        \node[right=of plot1] (plot2) {\includegraphics[width=0.4\linewidth]{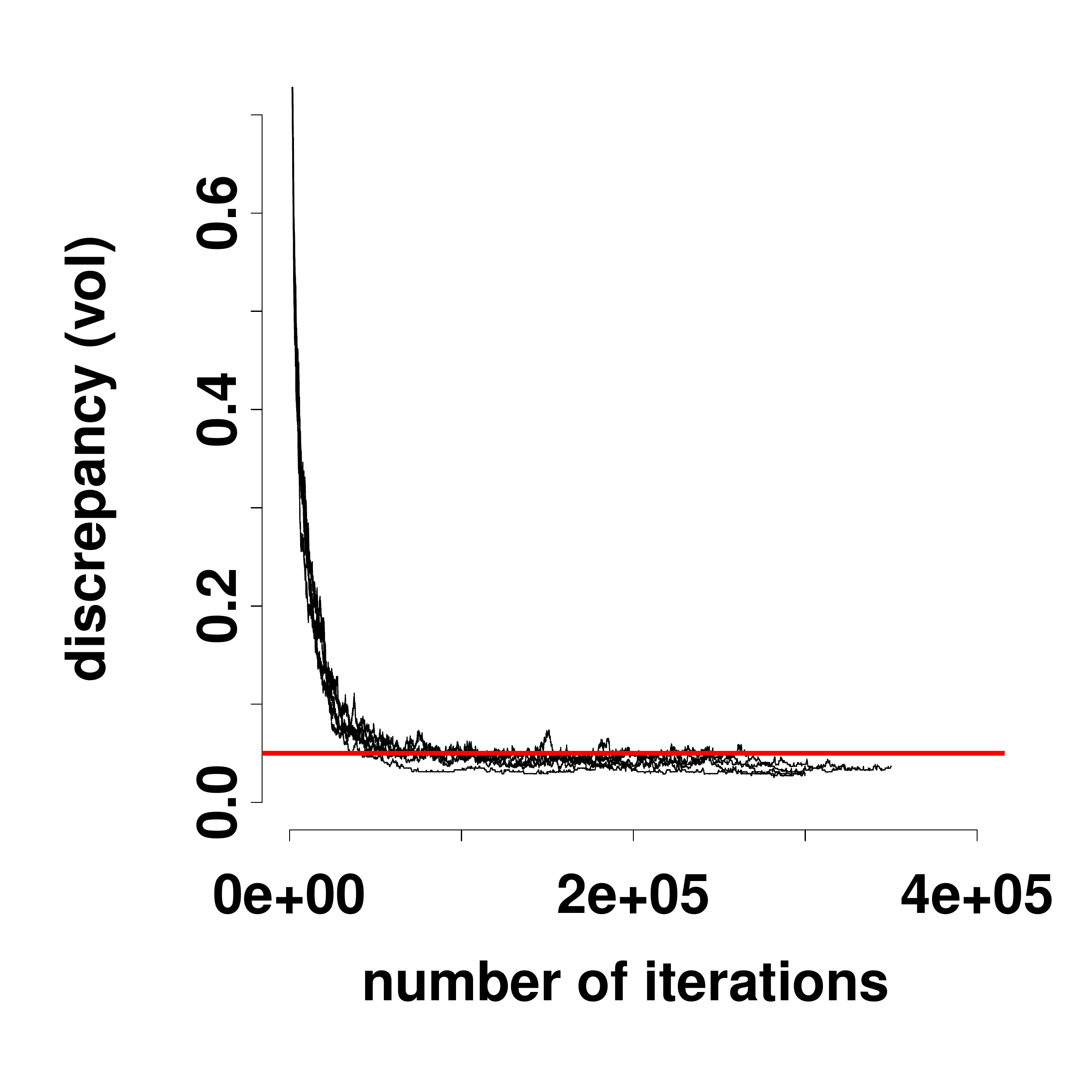}};

        \node at ($(plot1.north)$) {a)};
        \node at ($(plot2.north)$) {b)};
    \end{tikzpicture}
    \caption{Evolution of the discrepancy of histograms of cell volume for (a) the greedy algorthm and (b) the MHBDM algorithm. The red line represents the discrepancy $y=0.05$
}
    \label{vol_comp}
\end{figure}

\begin{figure}
\centering
    \begin{tikzpicture}[node distance=0cm]
        \node (plot1) {\includegraphics[width=0.3\linewidth]{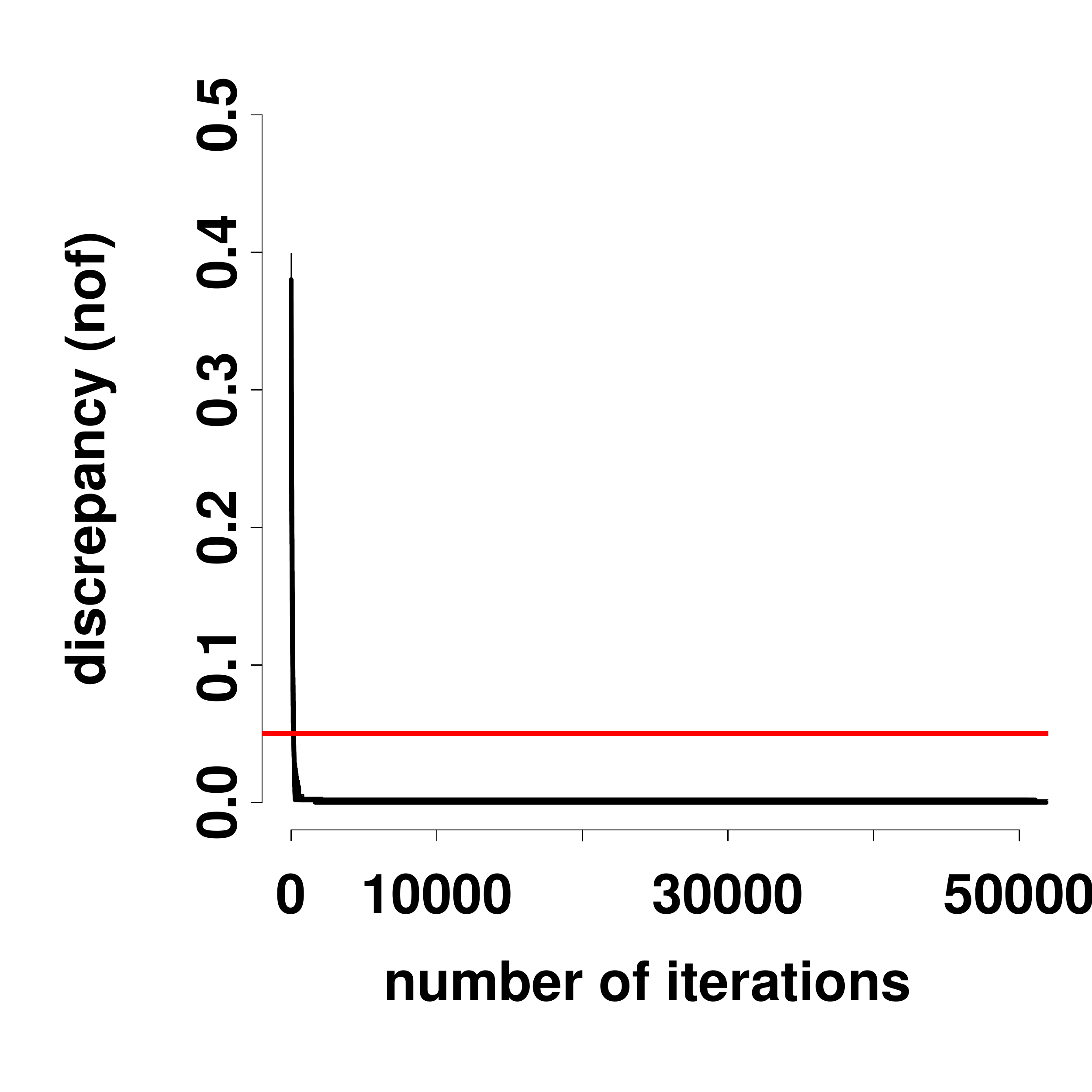}};
        \node[right=of plot1] (plot2) {\includegraphics[width=0.3\linewidth]{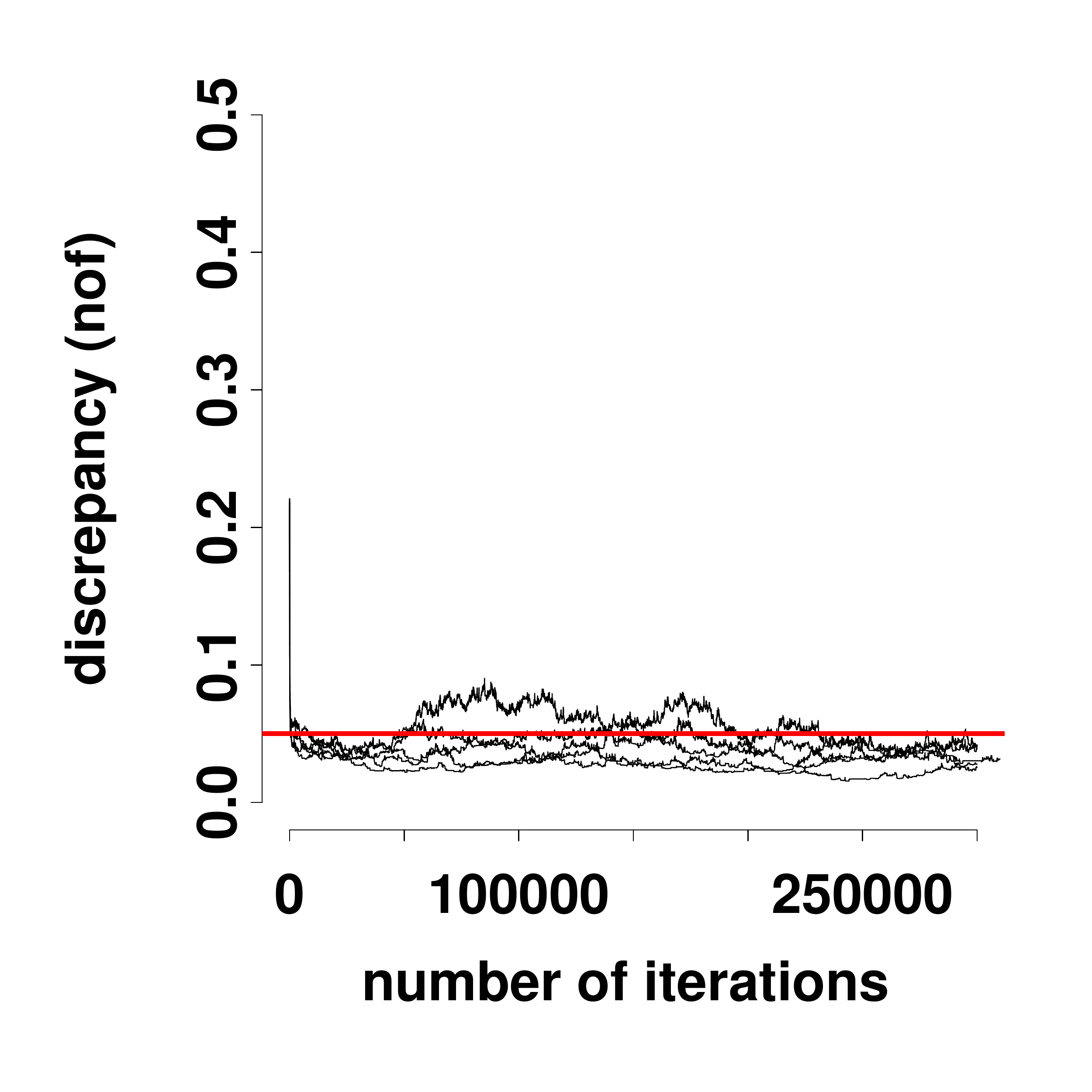}};
        \node[right=of plot2] (plot3) {\includegraphics[width=0.3\linewidth]{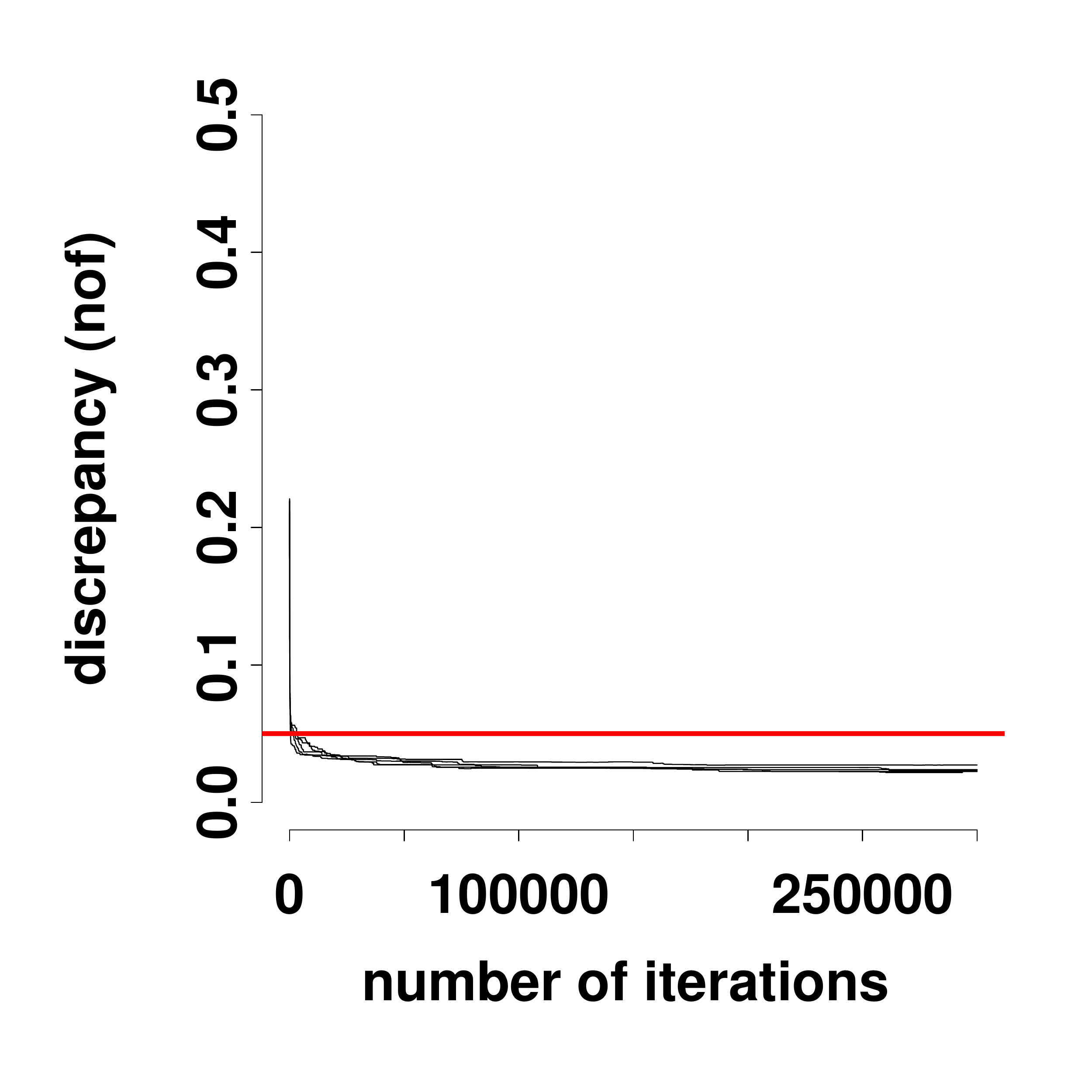}};

        \node at ($(plot1.north)$) {a)};
        \node at ($(plot2.north)$) {b)};
        \node at ($(plot3.north)$) {c)};
    \end{tikzpicture}

    \caption{Evolution of the discrepancy of histograms of number of faces per cell for (a) the greedy algorithm, (b) the MHBDM algorithm with $\theta_n = 1 \ 000$ and (c) the MHBDM algorithm with $\theta_n = 10 \ 000$.
     The red line represents the discrepancy $y=0.05$
}
    \label{nof_comp}
\end{figure}
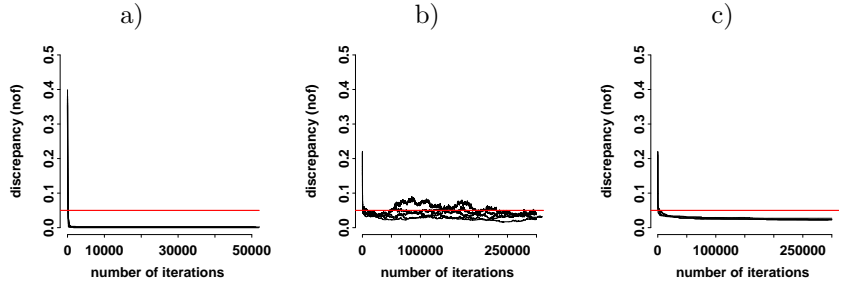

The reconstruction was carried out on the previously introduced experimental data set encompassing $1057$ nonempty cells. An important decision is which potentials will be incorporated in \eqref{energy_r}. We provide two comparisons, both defined by a single potential based on the histogram discrepancy and without any hardcore parameters. The first comparison concerns the volumes of cells. Both reconstruction approaches consider the discrepancy \eqref{dsc} between the histogram of cell volumes of each generated tessellation and the corresponding histogram of the experimental data $hist_{\vol}^{exd}$ shown in Fig.~\ref{stats}. In both algorithms there are some auxiliary parameters that need to be specified: namely, in \textit{Algorithm 3} we set $\theta_n = 1000$ and $(\delta,t) = (0.01,100000)$, and in \textit{Algorithm 4} we set $M=1057$ and $L=50000$.
In Fig.~\ref{vol_comp} we observe that the discrepancy stops decreasing after $200\,000$ iterations in the case of the greedy reconstruction and after $50\,000$ iterations in the case of the MHBDM reconstruction. The computational time to arrive at this point is roughly the same for both approaches.
It seems that the greedy reconstruction has the natural advantage that the mean cell volume is guaranteed to be the correct value throughout the entire run. Despite this fact, the MHBDM reconstruction yields smaller discrepancies, as can be seen in Fig.~\ref{vol_comp}. On the other hand, when dealing with histograms of the number of faces per cell, the results, cf. Fig.~\ref{nof_comp}, are better for the greedy reconstruction.

In the literature, cf. e.g., \cite{Il}, statistical reconstruction of point patterns is considered to be a non-parametric method. Our method interconnects statistical reconstruction with the simulation of stationary Gibbs point processes and uses auxiliary parameters to control the precision of the fit. Altogether, there is a common step, \textit{Algorithm 1}, that can be used in the simulation of marked Gibbs point processes and the reconstruction of marked point patterns. Note that, in contrast to the classical reconstruction \cite{Il}, the number of points in the reconstructed pattern does not need to be fixed. 
In summary, the benefit of the MHBDM reconstruction compared to the greedy algorithm introduced in \cite{CL} is that the former allows more flexibility in how close the reconstructed tessellation tracks the data. Moreover, since the MHBDM algorithm is not prone to getting stuck in local minima, better fit can be achieved.




\section{Concluding remarks}\label{con}
The Gibbs-Laguerre tessellation 
is a much more flexible stochastic model than the Poisson-Laguerre and Gibbs-Voronoi tessellations previously studied in the literature. This is demonstrated by several numerical studies in the three-dimensional Euclidean space. A certain disadvantage is the fact that conventional methods for parameter estimation--which work well for Gibbs particle systems [3]--give satisfactory results in the case of Gibbs-type tessellations only for small ranges of the activity and parameters $\theta$.
Therefore, we focus more on the 3D statistical reconstruction of tessellations 
derived from experimental image data
from materials research, extending some earlier approaches to Gibbs-type tessellations. By constructing the energy function in an appropriate manner, we are able to control the geometrical characteristics of interest, obtaining tessellations that are  comparable with a given data specimen. The development of such methods is important for the generation of virtual polycrystalline microstructures, 
whose physical properties can be investigated by means of numerical modeling and simulation. 
It is shown that by using MCMC techniques, we are able to generate different realizations of tessellations that provide  good reconstructions in the statistical sense.  This was verified for characteristics like the number of faces per cell and the cell volume. Alternatively, a number of other characteristics can be used instead, e.g., the NVR statistic, which introduces spatial interactions of neighboring cells.

\section*{Acknowledgements}
The authors acknowledge the financial support for this work provided by the Czech Science Foundation, project 17-00393J and by the German Research Foundation (DFG), project SCHM997/23-1.

\bibliographystyle{spmpsci}      
\bibliography{biblio}   

\end{document}